\documentclass[10pt,journal,compsoc]{IEEEtran}

\usepackage{amsmath}
\usepackage[T1]{fontenc}
\usepackage[latin9]{inputenc}
\usepackage{units}
\usepackage{graphicx}
\usepackage{esint}
\usepackage{booktabs}
\usepackage[]{cite}
\usepackage{balance}
\usepackage[caption=false,font=footnotesize]{subfig}
\usepackage{tikz}
\usepackage{enumitem}

\begin{document}


\title{Field-Programmable Crossbar Array (FPCA)\\for Reconfigurable Computing}

\author{Mohammed~A.~Zidan, YeonJoo~Jeong, Jong~Hoon~Shin, Chao~Du,
       Zhengya~Zhang,~\IEEEmembership{Member,~IEEE,}
        and~Wei~D.~Lu,~\IEEEmembership{Senior~Member,~IEEE}%
\thanks{The authors are with the department of Electrical Engineering \& Computer Science, University of Michigan, Ann Arbor, MI~48109, USA. e-mail: (wluee@eecs.umich.edu).}
}

\IEEEtitleabstractindextext{
\begin{abstract}
For decades, advances in electronics were directly driven by the scaling of CMOS transistors according to Moore's law. However, both the CMOS scaling and the classical computer architecture are approaching fundamental and practical limits, and new computing architectures based on emerging devices, such as resistive random-access memory (RRAM) devices, are expected to sustain the exponential growth of computing capability. Here we propose a novel memory-centric, reconfigurable, general purpose computing platform that is capable of handling the explosive amount of data in a fast and energy-efficient manner. The proposed computing architecture is based on a uniform, physical, resistive, memory-centric fabric that can be optimally reconfigured and utilized to perform different computing and data storage tasks in a massively parallel approach. The system can be tailored to achieve maximal energy efficiency based on the data flow by dynamically allocating the basic computing fabric for storage, arithmetic, and analog computing including neuromorphic computing tasks.

\end{abstract}

\begin{IEEEkeywords}
Cognitive Computing, Crossbar, Memristor, non-Von Neumann, RRAM
\end{IEEEkeywords}
}

\maketitle
\IEEEdisplaynontitleabstractindextext

\IEEEpeerreviewmaketitle


\IEEEraisesectionheading{\section{Introduction}}

\IEEEPARstart{T}{he} development of ever more powerful computing systems has primarily been driven by technology advances. Currently, billions of digital microprocessors play critical roles in our daily lives and empower our imaginations for a better future. However, modern computing tasks such as big data analysis, artificial intelligence, and pervasive sensing require energy efficient computing that cannot be fulfilled by the existing computing technology~\cite{src_2015}. For more than forty years, improvement in computer performance has been enabled by scaling down of CMOS transistors. This performance improvement slowed down after hitting the heat wall and memory wall, respectively~\cite{borkar2011future, darpa_2008, nair2015active}, and is approaching its physical scaling limits by the mid of 2020's~\cite{Shalf2015, waldrop2016chips}. Therefore, there is an urgent need to shift to new technologies, at both architecture and device levels where new physical phenomena and state variables can be used to store and process information. One such example is resistive random access memory, theoretically categorized as memristive devices or memristors~\cite{chua1971memristor, strukov2008missing}, which has attracted growing attention as a promising candidate for future data storage and computing due to its fast-operating speed, low power, high endurance, and very high density~\cite{itrs2,kim2010nanoscale,lee2011fast}.

Along its history, digital computers have passed through four generations, namely, Cathode Ray Tubes (CRTs), transistors, and Integrated Circuit (ICs)/microprocessors. Here it is clearly noted that the transition from one generation to the next is always marked by a technology advance at the device level. It is thus reasonable to expect that the recent advances in emerging device technologies~\cite{yang2012observation} may usher in a new computing era. For instance, the high-density memristor crossbar structure is widely considered one of the best candidates for nonvolatile storage and Random Access Memory (RAM) applications~\cite{kim2011functional, zidan2014memristor, wong2012metal, akinaga2010resistive, vontobel2009writing, zidan2016}. Furthermore, analog resistive devices have been shown to be well suited for bio-inspired neuromorphic computing systems~\cite{yang2013memristive, jo2010nanoscale, sheridan2015feature, prezioso2015training} and can significantly outperform classical digital computing in many ``soft'' computing applications where the task is complex but approximate solutions are tolerated, with such examples including data classification, recognition, and analytics~\cite{Shalf2015, alibart2013pattern, kim2015640m}. At the other end of the spectrum, many studies have been attempted to perform accurate digital computations using binary resistive memory devices~\cite{snider2005computing, likharev2006cmol, borghetti2010memristive, xia2009memristor}. In both cases, systems based on these emerging devices are normally used as accelerators for a subset of specialized tasks, e.g. data storage, neuromorphic computing, and arithmetic analysis, and each task uses different physical devices, circuits, and system organizations to achieve a specialized goal. While utilizing these subsystems in a traditional computing platform is expected to achieve improved performance, particularly for the target tasks, a general computing system that can handle different tasks based on a uniform physical fabric in a fast and energy-efficient manner is desired.

\begin{figure}[!t]
\vspace{-5pt}
\begin{centering}
\includegraphics[width=\columnwidth]{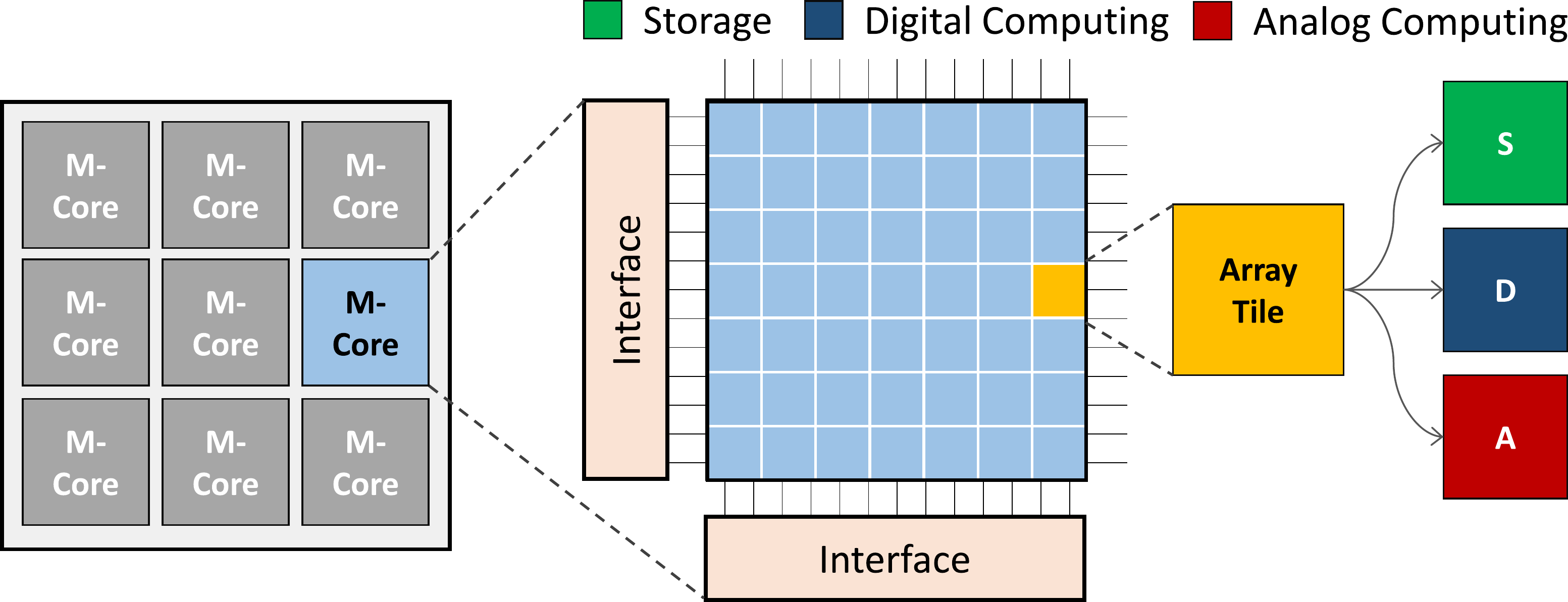}
\par\end{centering}
\caption{Block diagram showing the different layers of the proposed FPCA computing architecture.\label{fig:syst}}
\vspace{-17pt}
\end{figure}

\begin{figure*}[!t]
\vspace{-3pt}
\begin{centering}
\subfloat[\label{fig:config}]{\includegraphics[height=97pt]{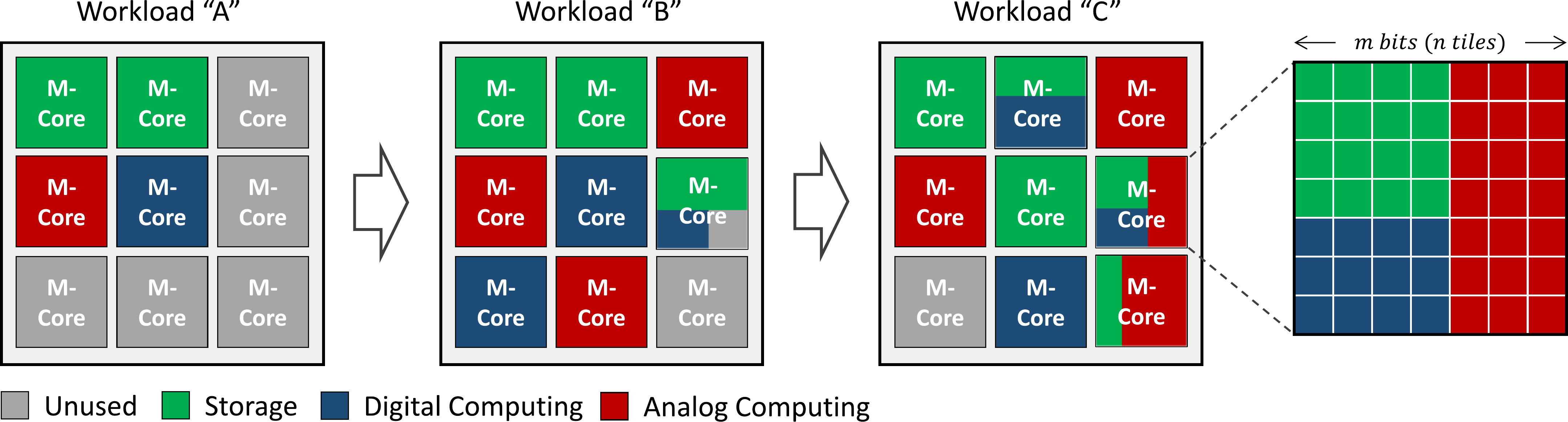}}\,\,\,\,\,\,\,\,
\subfloat[\label{fig:3d}]{\includegraphics[height=93pt]{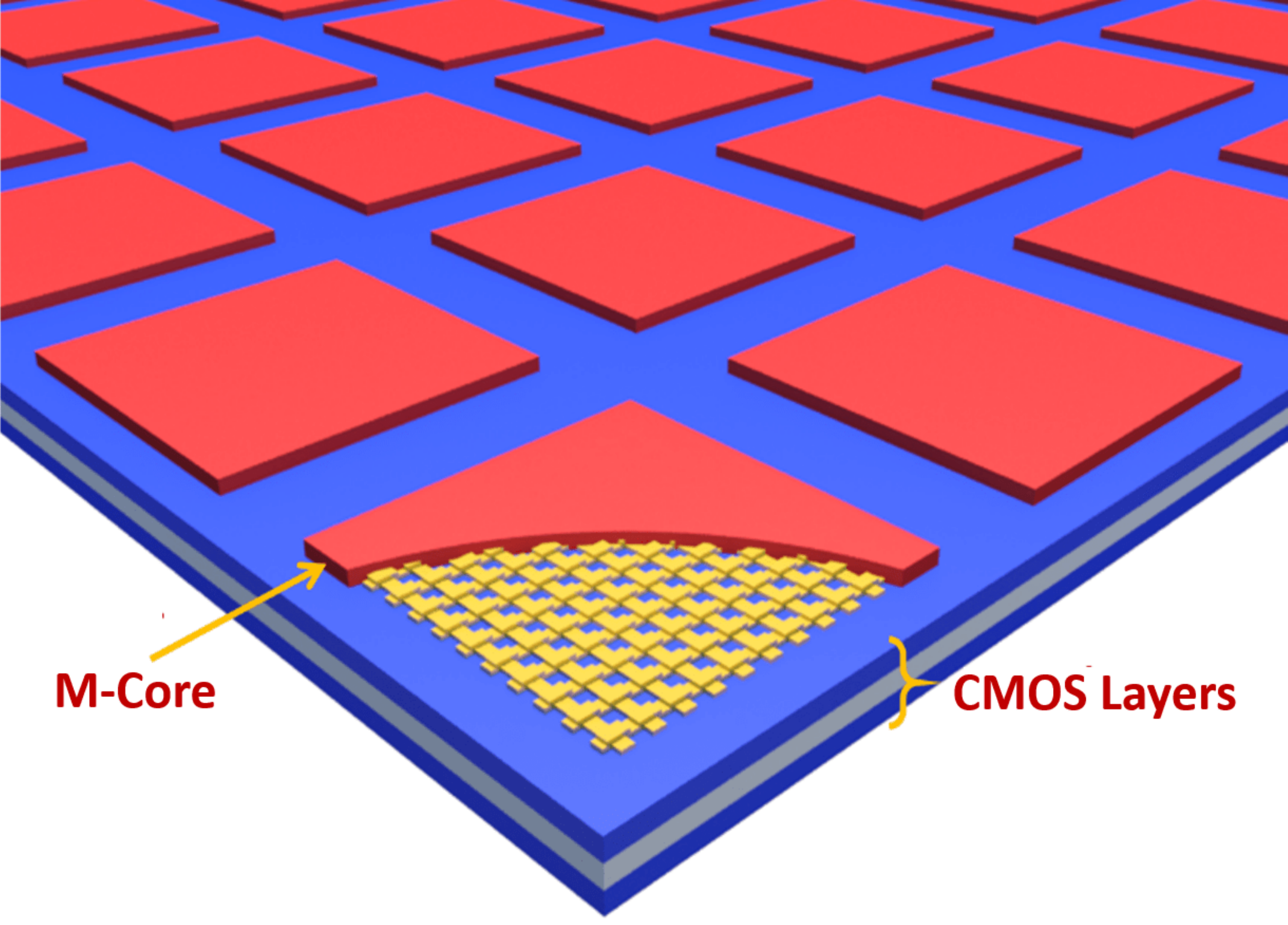}}
\par\end{centering}
\caption{(a) Different configurations for an FPCA system based on different computing workloads. (b) 3D illustration showing the M-Cores monolithically fabricated over the CMOS layers.}
\vspace{-10pt}
\end{figure*}

We believe that the optimal solution is to merge the three tasks, memory, analog computing and digital computing, together using a single physical fabric to achieve a general computing platform. In general, the memory wall needs to be overcome~\cite{borkar2011future, darpa_2008} by reducing the amount of slow and power-hungry communications between the memory and the processor. Moreover, computing methodology should be natively parallel at the fine grain level. Finally, it is desirable for a new computing architecture to incorporate analog computing capabilities to achieve better energy efficiency in tasks such as data analytics, classification, and recognition~\cite{Shalf2015}. We believe these requirements can be satisfied in a novel computing architecture which we term Field Programmable Crossbar Array (FPCA). The proposed architecture is built around the idea of having a universal core block that can be dynamically reconfigured to serve different workloads optimally, schematically shown in Figure~\ref{fig:syst}. In this approach, the resistive crossbar's inherent parallelism is optimally utilized at the physical device level to directly perform different computing and data storage operations efficiently, while at the system level the architecture can dynamically reallocate resources to optimally match the computing needs for the incoming data. The main challenge here is how to utilize a common physical fabric (the resistive crossbar and its interface circuitry) to perform the three sets of diverse tasks that typically require three completely different systems.

In this work, we show that the crossbar array based common physical block can indeed store data and process in-memory processing in analog and digital fashion. Utilizing binary resistive crossbar as the common physical block, we show the system can efficiently implement binary neural networks, arithmetic tree reduction, and in-situ data migration. These operations allow the proposed FPCA computing system to provide three important functions. Firstly, the ability to process any arbitrary workload in its optimal computing domain (digital or analog). Secondly, the natively modular design of the system allows a high degree of scalability and reconfigurability to tailor fit different workloads. Finally, it merges processing and memory together at the lowest physical level to achieve maximal efficiency and minimal data migration. Our analysis shows an FPCA-based high-performance computing system offers a much smaller energy budget compared to classical Von Neumann architectures in both classical and cognitive computing applications.


\section{FPCA Computing Architecture}
The proposed FPCA architecture is organized in a hierarchical array structure, where the top layer is composed of crossbar modules (Memory cores, M-Cores). Each M-Core is a single crossbar that can compute with/in local memory. Each M-Core is further (virtually) divided into a set of tiles. While all the tiles are physically identical, each of them can be dynamically re-configured to perform one of the three different tasks, storage (S), digital computing (D), or analog computing (A). Therefore, the system can offer different modes of operations at the fine grain level. As will be shown later, this approach enables natively scalable, reconfigurable and energy-efficient computing. Figure~\ref{fig:syst} shows a block diagram illustrating the different layers of the FPCA architecture, showing the M-cores at the system level and the individual tiles within each M-core.

The new computing system can be configured either at the system level or the core level. At the system level, an entire M-core is assigned to a particular task, for example, one core for analog computing. This core can be later reassigned to digital computing or used as storage based on computational need. Finer grain configuration can be achieved by assigning different tiles of a given core to different tasks. Such a low-level configuration is optimal for high throughput data processing and analysis, where the stored data can be processed by the same core in both digital and analog schemes, without the need to move the data back and forth between processing and storage cores. A more generic approach allows the resource reconfigurations on the two levels simultaneously based on the nature of the workload, as shown in  Figure~\ref{fig:config}. This configuration scheme is equivalent to having a pool of generic resources, where they are assigned to perform specific tasks based on the workload requirements. The system dynamically reconfigures to adapt to the workload. It should be noted that one of the essential characteristics of the proposed architecture is the resistive crossbar being natively modular, parallel, and reconfigurable. This allows the system to scale from a small scale IoT smart node chip to a supercomputing type of architecture.

Besides reconfigurability, another aspect in the design of the FPCA system is energy efficiency. It is challenging to implement energy efficient systems at different scales since there is no universal approach for energy efficient computing. For instance, small and medium computing systems require partial or fully sleep mode to achieve energy efficiency, as in smart nodes and mobile devices. FPCA achieves this by utilizing the nonvolatile property of its resistive memory devices, where the system can go to a zero-power sleep mode without the need to spend power to keep track of the system state. On the other hand, a large computing system requires an energy efficient data flow and parallel processing units, which already exist as the core properties of the FPCA architecture. Combined with the multi-domain computing capability where tasks can be processed in the native domain (either analog or digital), these features make the FPCA a very fast and energy efficient computing system.

\subsection{Reconfigurable M-Core}
A key property of the FPCA architecture is the ability of an M-core to be reconfigured to perform different tasks. Each M-core is composed of a crossbar array and its interface circuitry, as shown in Figure~\ref{fig:syst}. A major challenge of the FPCA architecture is to map different computing and storage tasks to the single common physical fabric, the M-core. This starts by selecting the right RRAM candidate. We found that binary RRAM devices are suitable candidates to implement the M-Cores that can perform the different operations required by FPCA. These devices are well-known for their very high density, low power consumption, and fast access speed~\cite{jo2009high, kozicki2005nanoscale}. Such outstanding properties make them attractive as a future replacement for Flash-based memory and storage, although their applications in computing is less explored compared to analog memristors. Below we show that the binary memristor devices can be optimally utilized for both digital and analog computing tasks, besides being used as data storage devices. With this approach, all three subsystems (storage, analog and digital computing) can be implemented using a common physical fabric to allow the computing tasks to be performed efficiently, as elaborated in the following sections.

\subsection{3D Monolithic Chip}
The FPCA system relies on recent advances in RRAM technology to provide the system with its computational and storage capabilities~\cite{itrs2,lee2011fast,gaba2014ultralow}. Only a small CMOS component is required to provide necessary peripheral functions such as interface and control circuitry. In this regard, the CMOS system can be considered as the accelerator component while the M-Cores perform the general computing tasks. We envision a monolithic approach to building a 3D computing chip, where the high-density memristor crossbar is fabricated on top of the CMOS circuitry as shown in Figure~\ref{fig:3d}. It has already been demonstrated that RRAM crossbar fabrication requires low thermal budget, and hence can be safely fabricated on top of a typical CMOS process~\cite{kim2011functional, shulaker2014monolithic, aly2015energy, rev2,rev1} for memory and in-memory digital computing applications. The monolithic integration allows distributed, local, and high-speed interface between the RRAM layer the and CMOS layer underneath. The CMOS layer will host the analog interface for the M-Cores, which includes analog MUXs and ADCs. This will allow a parallel access to a full tile per each M-Core. Additionally, the CMOS layer will host fast interconnect and other digital periphery circuitry. The CMOS/crossbar integration will likely follow earlier studies, where successful CMOS/RRAM hybrid systems have been demonstrated for memory applications~\cite{kim2011functional}.


\section{In-Place Arithmetic Operations}
Arithmetic operations are the foundation of any digital computational system, where the performance of digital computers is typically measured in FLOPS (floating point operations per second). Almost every arithmetic operation relies on a tree reduction circuit to perform functions such as multiplication, division, trigonometric operations, matrix operation and multi-operand addition. In tree reduction, multi-operand additions are transformed to two-operand additions. This seemingly simple task consumes most of the arithmetic units' area and energy budget. Typically tree reduction is realized using successive stages of arithmetic counters and compressors (a generalized form of full adders)~\cite{Flynn2001}. There are various flavors of the arithmetic trees, with clear tradeoffs between area and speed. However, all of them are built around the idea of cascading and looping over arithmetic compressor units. An arithmetic compressor counts the number of ONEs per input. For instance, an n-operand adder is just a group of cascaded arithmetic compressors.

Here, we propose to perform massively parallel arithmetic operations directly in an M-core, where the crossbar structure is utilized as a giant arithmetic compressor. In the presented technique, multiple tree reduction operations can be performed simultaneously on the same crossbar array. Moreover, masked tree reduction is also feasible, thus eliminating the need for extra logic gates for many of the arithmetic operations, e.g. in multiplications. These capabilities allow M-cores to perform in-memory parallel digital processing efficiently and natively.

\begin{figure}[!b]
\vspace{-15pt}
\begin{centering}
\subfloat[\label{fig:cb_activation_a}]{\includegraphics[height=111pt]{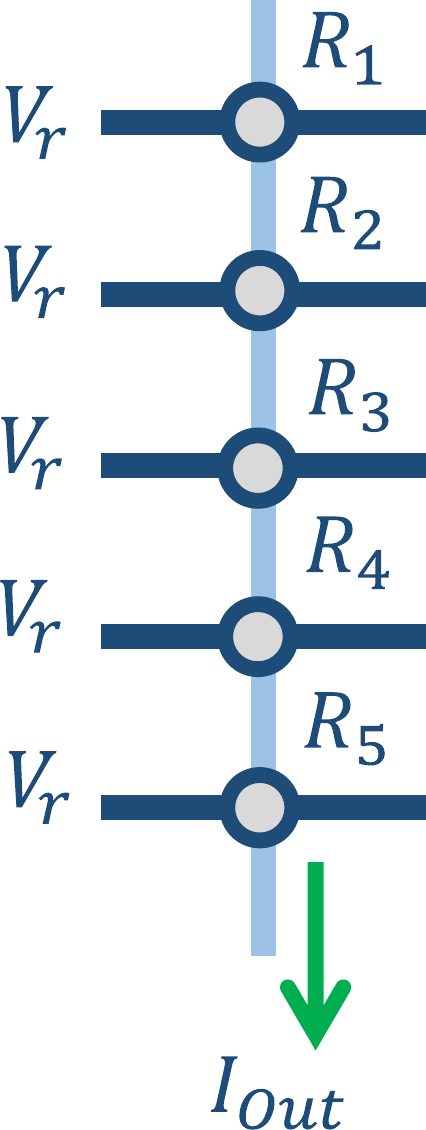}}\,\,\,\,\,\,\,\,\,\,
\subfloat[\label{fig:cb_activation_b}]{\includegraphics[height=111pt]{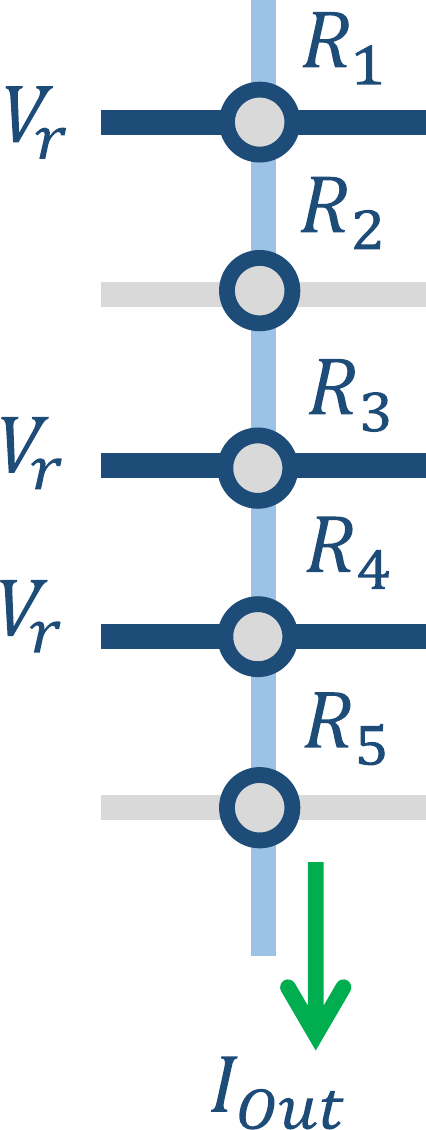}}\,\,\,\,\,\,\,\,\,\,
\subfloat[\label{fig:cb_activation_c}]{\includegraphics[height=111pt]{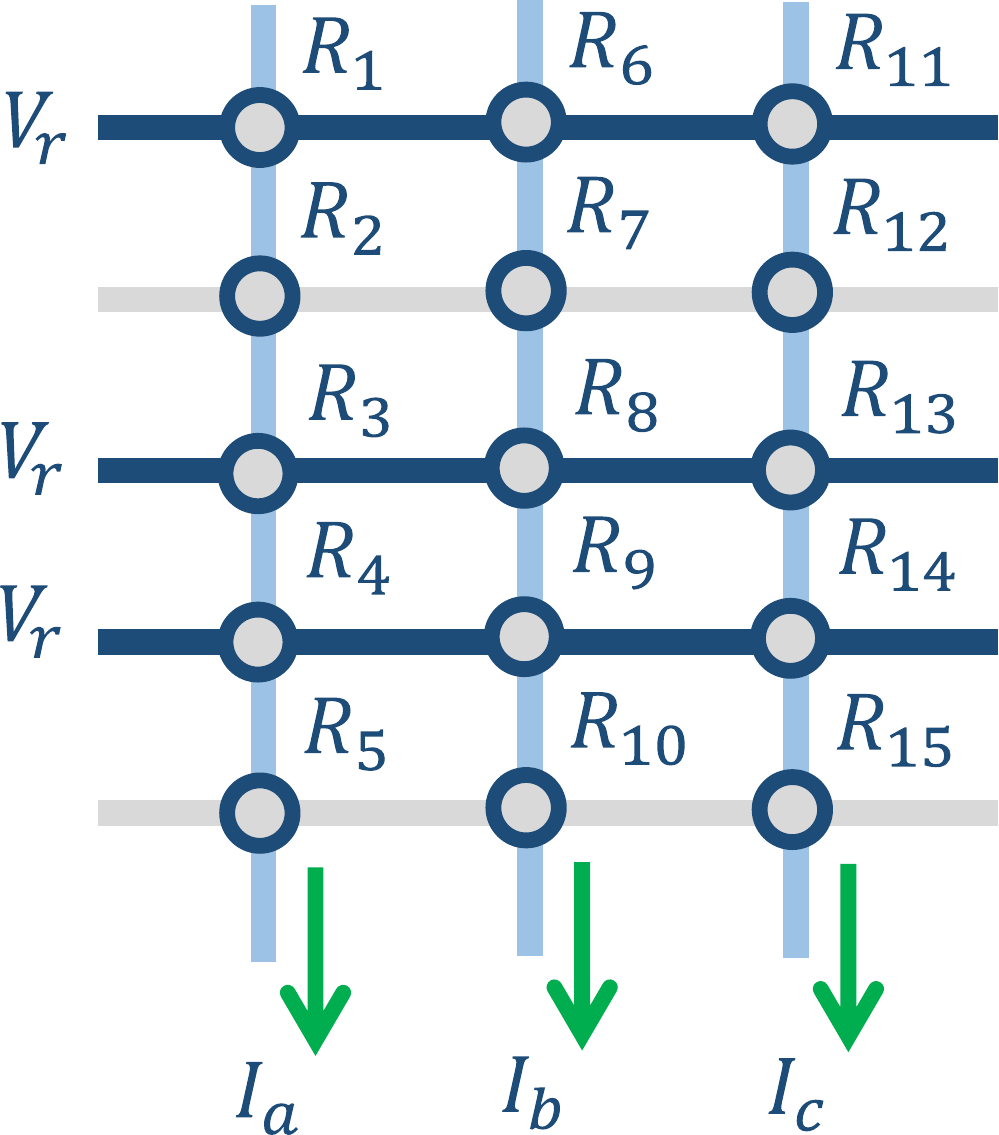}}
\par
\end{centering}
\caption{Unmasked and masked crossbar activation.\label{fig:cb_activation}}
\end{figure}

\subsection{Counting the Ones}
The basic concept of any arithmetic compressor is to count the number of ONEs, and this can be achieved efficiently in a crossbar structure. We first examine a single column inside a crossbar, with all its rows biased with a reading voltage, as shown in Figure~\ref{fig:cb_activation_a}. The output current is described as,
\begin{equation}
I_{out}=V_{r}\sum\frac{1}{R_{i}}
\end{equation}
Knowing that $R_i=\{R_{on},R_{off}\}$ and $R_{off}\gg R_{on}$, the output current can be rewritten as, 
\begin{equation}
I_{out}\approx N_{ones}\left(\frac{V_{r}}{R_{on}}\right)
\end{equation}
where ``$N_{ones}$'' is the number of ONEs in the column, and ``$V_r/R_{on}$'' is a constant value. The read current can then be readily translated into a digitized value with the aid of the common interface circuitry of the M-core, where the interface circuit digitizes the crossbar readout current into binary bits with the aid of the ADCs, where the same ADC circuitry will be utilized for different types of M-core's operations. A masked version of the tree reduction can be achieved by only biasing the rows of interest, as shown in Figure~\ref{fig:cb_activation_b}. This significantly simplifies multiplication and division operations by eliminating the need for AND gates. In such case, the output current is written as
\begin{equation}
I_{out}=\frac{V_{1}}{R_{1}}+0+\frac{V_{3}}{R_{3}}+\frac{V_{4}}{R_{4}}+0+\ldots
\end{equation}
which is equivalent to the following summation,
\begin{equation}
S=A\wedge W+B\wedge X+C\wedge Y+D\wedge Z+\ldots
\end{equation}
where the equation is written using dummy variables. The simple circuit realization of this equation is the key to the crossbar based arithmetic calculations. The masked tree reduction can be further extended to multiple columns in a natively parallel fashion, as shown in Figure~\ref{fig:cb_activation_c}.

\begin{figure}[!t]
\vspace{-7pt}
\begin{centering}
\subfloat[\label{fig:ideal}]{\includegraphics[width=.83\columnwidth]{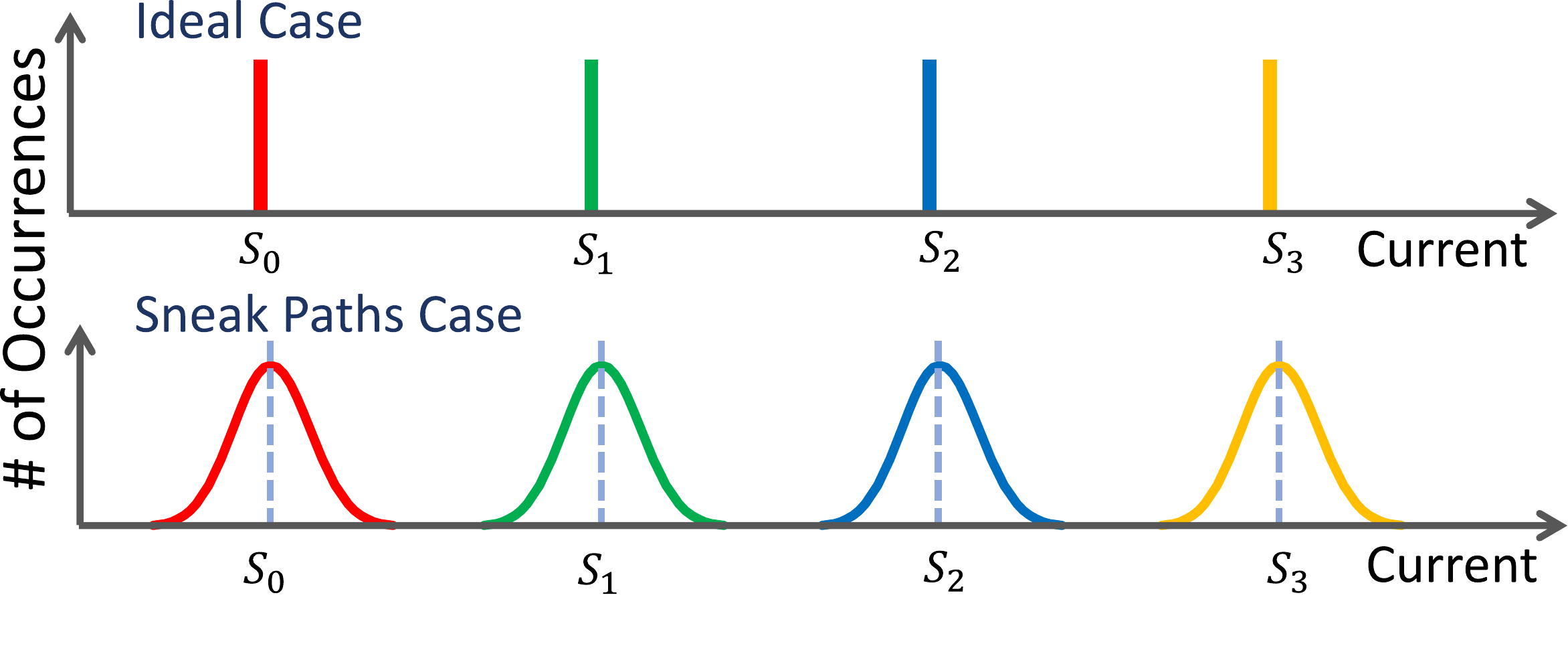}}\\
\subfloat[\label{fig:staircase}]{\includegraphics[width=0.71\columnwidth]{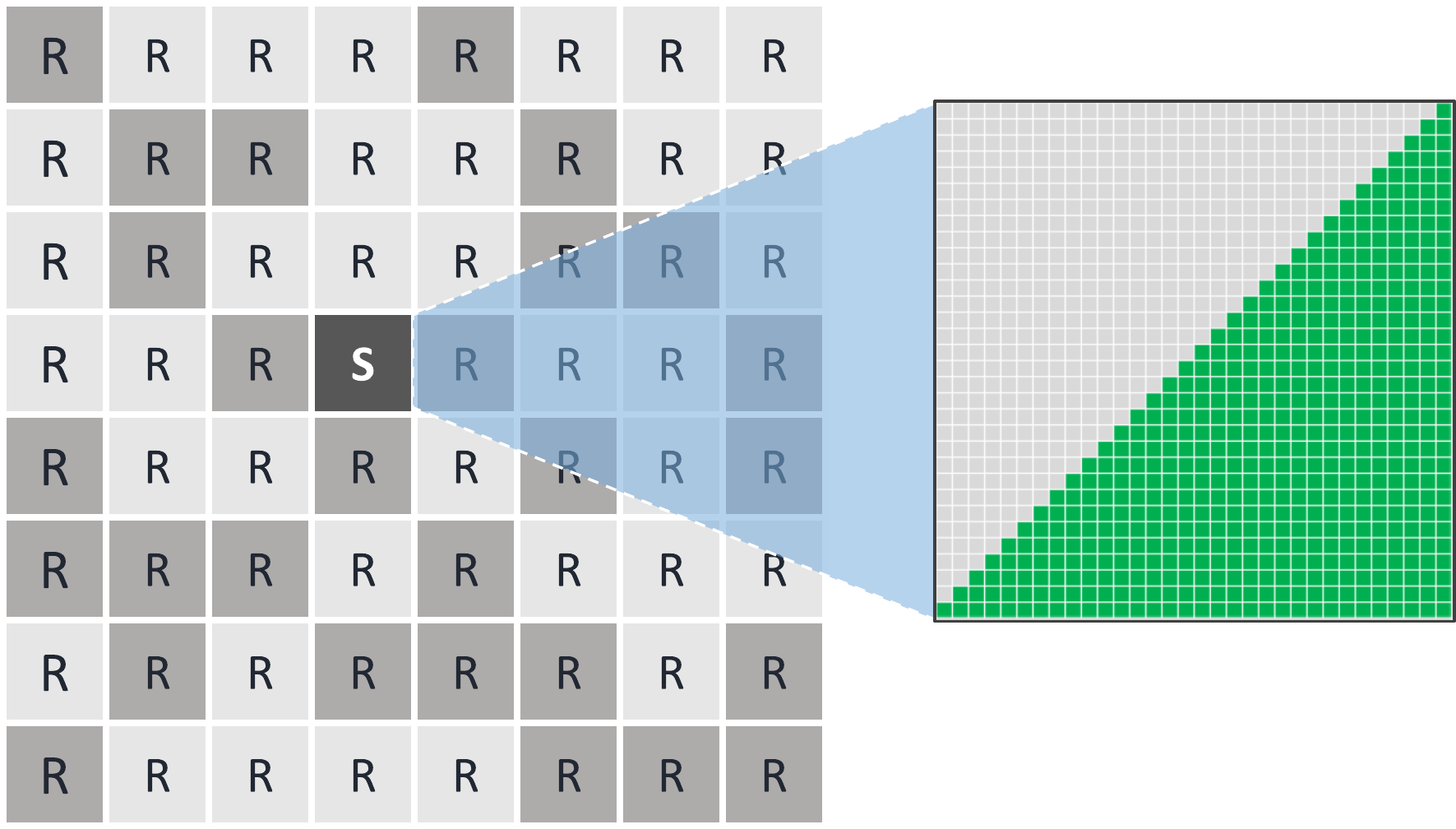}}
\par
\end{centering}
\caption{(a) The ideal and the practical cases for a column readout currents in the absence and presence of sneak paths. (b) A sub array with all its tiles filled with random data patterns, except for the target tile which is filled with staircase like data to verify the ability of counting the number of ONEs per a tile column.}
\vspace{-10pt}
\end{figure}

The data stored in a column of n-bits can represent (n+1) different symbols depending on the number of ONEs per column. During a full column activation, each symbol should have a distinguishable current level. However, since the current flows through the rest of the crossbar cells, each symbol is now represented by a distribution rather than a single value as shown in Figure~\ref{fig:ideal}. We need to design our system to properly differentiate different symbols and compensate any undesired effects. Hence, we built an accurate Python/HSPICE simulation platform to simulate the proposed FPCA arrays. The platform is designed to simulate the different modes of the FPCA operation for any arbitrary set of data. Moreover, it also accounts for the different biasing and connectivity schemes. The simulation platform adopts experimental device models and accounts for crossbar parasitic nonidealities, such as the crossbar line resistance and the switching circuitry. These usually overlooked parasitic effects can potentially significantly alter the simulation results as discussed in~\cite{zidan2014memristor}. Figure~\ref{fig:staircase} shows an M-core consisting of 256 tiles, each of which is in turn 1k bits (32x32) in size. One of the tiles is filled with a staircase pattern with an increasing number of ONEs per column. All the other tiles are filled with random data, and the system is simulated with more than 44k different data patterns. The purpose of these simulations is to verify the M-core's ability to count the number of ONEs correctly despite the unknown content of the surrounding tiles and parasitic effects such as the sneak paths. During operation, all rows and columns of the tile of interest are activated simultaneously so that the number of ONEs per column for all the tile columns can be read out in one step. Besides increasing the degree of parallelism, this full tile access approach significantly reduces the sneak paths effect. Finally, it should be noted here that the RRAM device ON/OFF ratio needs to be much higher than the number of active rows so that the sum of the ZEROs is not misclassified as ONE. Luckily, ON/OFF ratio of > 32 or 64 are readily achievable in binary RRAM devices.

\begin{figure*}[!t]
\begin{centering}
\subfloat[Grounded terminals]{\includegraphics[width=0.65\columnwidth]{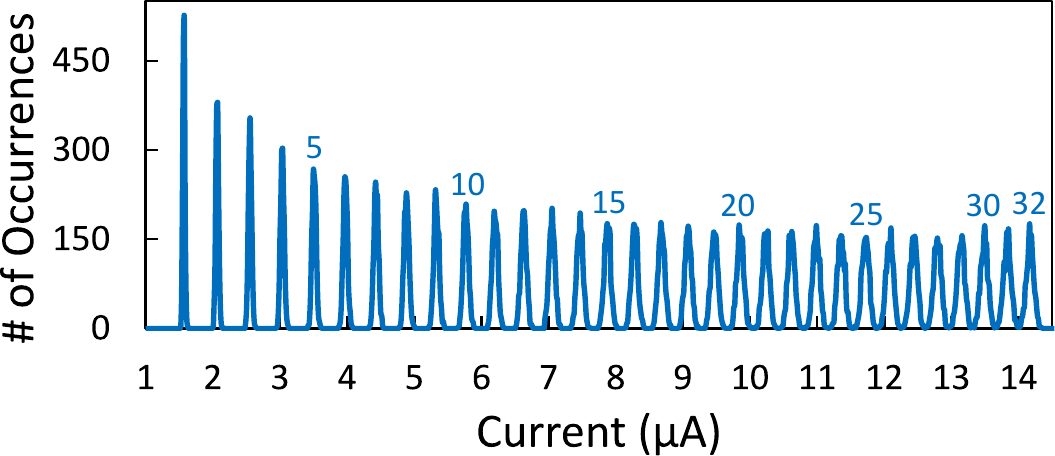}}\,\,\,\,\,
\subfloat[Half-selected terminals]{\includegraphics[width=0.65\columnwidth]{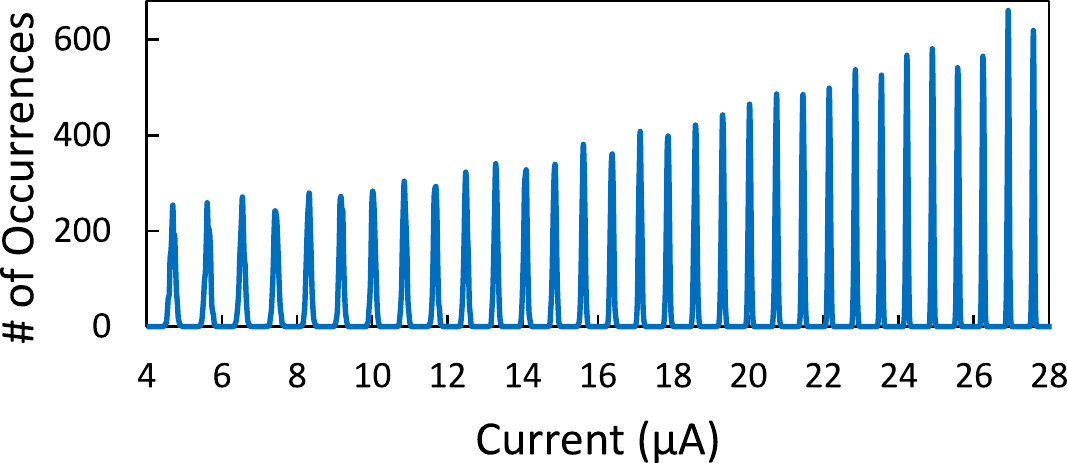}}\,\,\,\,\,
\subfloat[Floating Terminals]{\includegraphics[width=0.653\columnwidth]{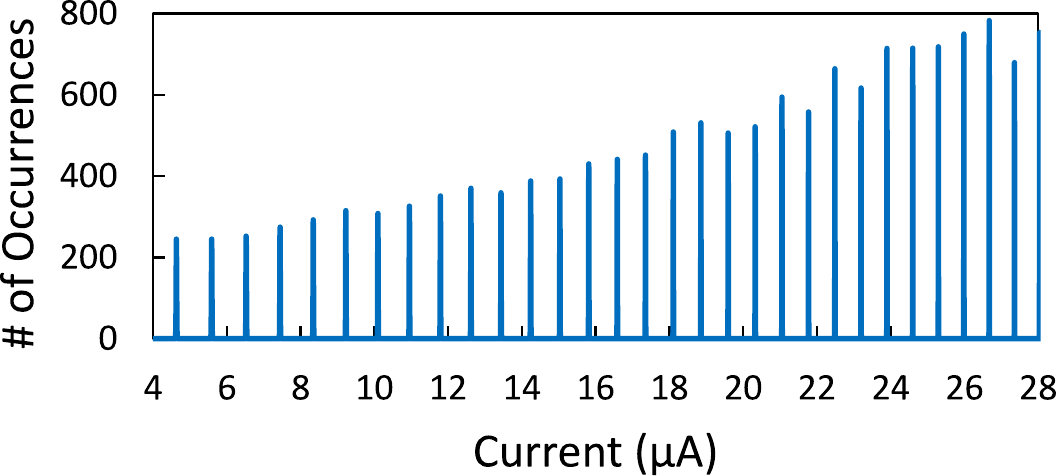}}
\par
\end{centering}
\caption{Histogram for the current readout from 32 different columns of a given tile, from 44,800 simulations points, where the rest of the M-Core is filled with random data.\label{fig:hist}}
\vspace{-10pt}
\end{figure*}

Figure~\ref{fig:hist} shows the simulation results as a histogram distribution of different output current levels, where each current level indicates a different number of ONEs. The results show that the center of the output distributions are equally spaced from each other, where each step in the current is equivalent to an extra ONE in the column count. The system is simulated multiple times with different techniques for connecting the unselected rows and columns. It turns out that grounding the unselected rows and columns leads to more smeared (but still separable) output patterns, and keeping the unselected rows and columns floating leads to better outputs, as shown in~Figure~\ref{fig:hist}. This is because grounding the rows and columns encourages the current to sneak out of its desired path. Hence, the measured current at the columns of interest will depend on the data pattern in the unselected tiles. On the other hand, floating unselected rows and columns effectively utilizes the high nonlinearity of the RRAM device to suppress sneak current. This effect is clearly visible in Figure~\ref{fig:hist}, where the current spread is minimal, and the separation is maximized. Grounding unselected rows and columns also increases the total power consumption because of the parasitic current component, but this approach may be more preferable from a circuit designer's point of view. The total power consumption for counting the number of ONEs in a given tile is 4.33mW, 1.1mW, and 1.06mW for grounded, half-selected, and floating terminals connection schemes respectively, where the RRAM device presented in~\cite{wang2015conduction} has been used in theses simulations.

\subsection{Arithmetic operations}

\begin{figure}[!b]
\vspace{-10pt}
\begin{centering}
\includegraphics[width=\columnwidth]{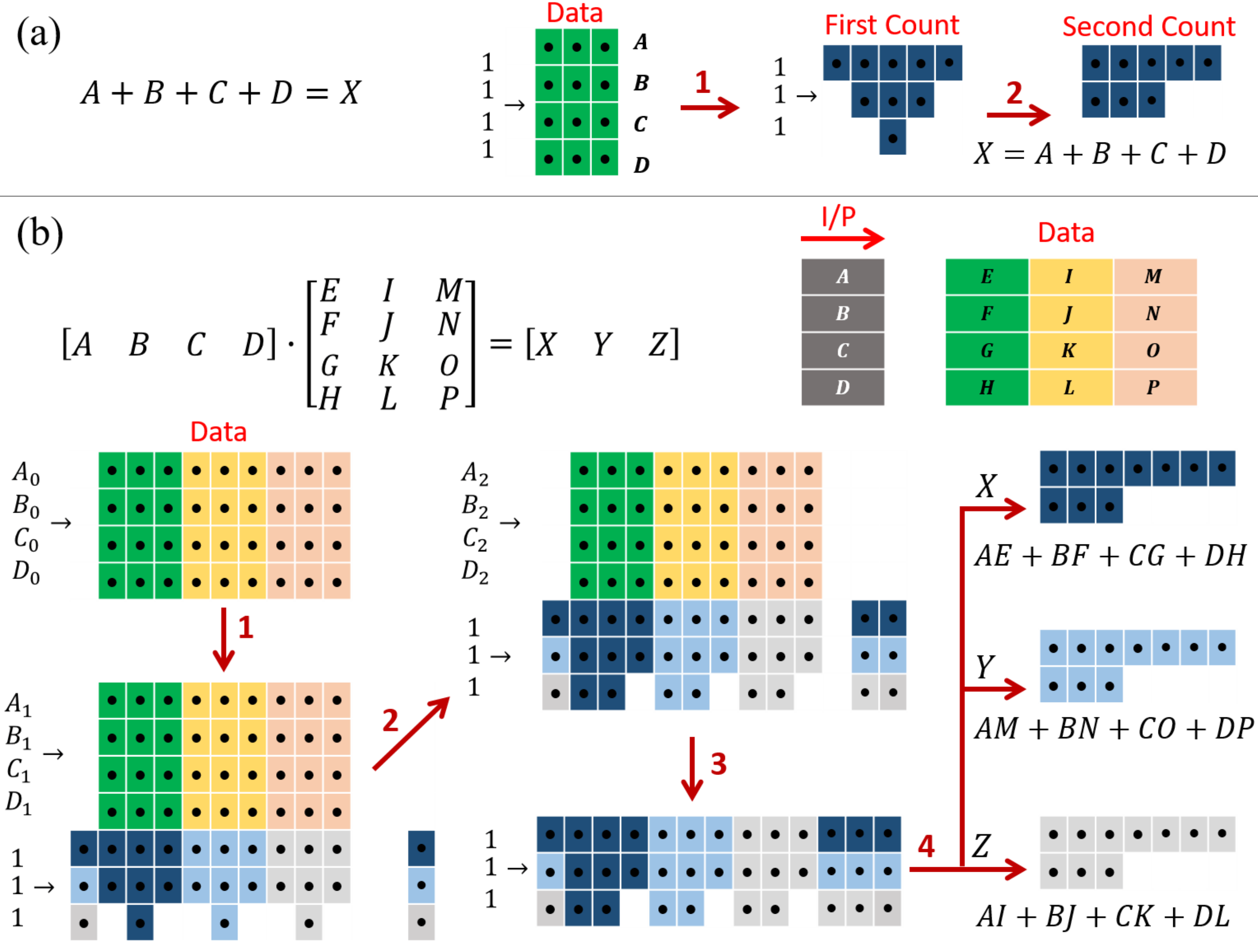}
\par\end{centering}
\caption{(a) Parallel vector addition and (b) Parallel vector-matrix multiplication steps using an M-core.\label{fig:mul}}
\vspace{-5pt}
\end{figure}

The ability of the M-cores to perform in-memory parallel tree reduction enables the implementation of different types of arithmetic operations. The simplest operation that can be implemented using the unmasked ONE-counting approach is parallel vector addition. In this case, the output of each column, which is the number of ONEs it contains, is written back to the M-core for the next operation. This process is repeated iteratively until the vector operation is reduced to a simple 2-operand addition, as shown in Figure~\ref{fig:mul}a. The parallel addition can then be extended to a more complex operation with the aid of masked tree reduction. For example, a multiplication operation is typically implemented using a tree adder, where the adder inputs are the different bits of the multiplicand and the multiplier added together. This can be illustrated in the following example showing a 3-bit operands multiplication:
{\small \begin{multline}
A\cdot B= +  \begin{array}{ccccc}
 &  & B_{0}A_{2} & B_{0}A_{1} & B_{0}A_{0}\\
  & B_{1}A_{2} & B_{1}A_{1} & B_{1}A_{0}\\
B_{2}A_{2} & B_{2}A_{1} & B_{2}A_{0}
\end{array}
\end{multline}}
Both the tree addition and the AND operation are performed using masked tree reduction. The multiplication process can be further extended to a dot-product operation. This vector operation follows the same structure of the basic multiplication operation, as shown in the following 3-bit dot product example:
{\small \begin{multline}
\left[A,B\right]\cdot\left[\begin{array}{c}
C\\
D
\end{array}\right]= \begin{array}{ccccc}
 &  & A_{o}C_{2} & A_{o}C_{1} & A_{o}C_{o}\\
 &  & B_{o}D_{2} & B_{o}D_{1} & B_{o}D_{o}\\
 + & A_{1}C_{2} & A_{1}C_{1} & A_{1}C_{o}\\
 & B_{1}D_{2} & B_{1}D_{1} & B_{1}D_{o}\\
A_{2}C_{2} & A_{2}C_{1} & A_{2}C_{o}\\
B_{2}D_{2} & B_{2}D_{1} & B_{2}D_{o}
\end{array}
\end{multline}}
Here, we need to implement this vector dot product operation using the masked tree reduction with minimal data movement. This can be implemented using the following proposed vector-vector multiplication algorithm. Let's call the first vector ``input vector'' and the second vector ``data vector''. The data vector will remain in its storage tile, while the input vector values will be used to activate the tile's row inputs. The data vector is organized such that its elements are arranged in a stacked form. The vector-vector multiplication algorithm is given below:
\begin{enumerate} [label=(\alph*)]
\item Use the first bit of all the elements of the input vector to activate the rows, where the read voltage is applied to a row in case of ONE; otherwise, the row is kept floating. 
\item Digitize the readout current of all the columns of interest, where the columns current is proportional to the number of ONEs per column within the activated row region. 
\item Write the counting output, shift one bit to the right, below the data vector, which we call compressed rows. 
\item Repeat steps ``a'' to ``c'' for the whole multiplier vector width. 
\item Apply read voltage to the compressed data rows. 
\item Digitize the readout current of all the columns of interest. 
\item Overwrite the compressed data with the new iteration results. 
\item Repeat ``e'' to ``g'' steps until a two-operand addition case is reached. 
\end{enumerate}
This algorithm can be extended to a vector-matrix multiplication as illustrated in Figure~\ref{fig:mul}b, where the vector-matrix multiplication can be implemented in parallel by activating all the columns and thus requires the same number of steps as a vector-vector multiplication. Using the same scheme, matrix-matrix operation can be performed in the crossbar structure. The proposed strategy applies to any tree-reduction based arithmetic operation, that is, typically any arithmetic operation other than incrementing or two operand addition. It can also account for signed operations with the aid of sign extensions. Finally, it should be noted that the final output of the tree reduction is always a 2-operand addition, which can be performed sequentially on the crossbar or a simple 2-operand adder in the system's CMOS layer.

\section{Binary Coded Neural Networks (BCNN)}
Another important aspect of the proposed architecture is the implementation of neuromorphic computing. This approach is generally inspired by how the biological brain processes data, where neural networks are used to execute complex operations in parallel. Such a computational technique can be extremely power efficient when processing cognitive applications compared to classical processors~\cite{merolla2014million}. Previous studies have shown that high-density (analog) memristive crossbar is one of the best candidates for realizing synaptic meshes in neural networks~\cite{ jo2010nanoscale, sheridan2015feature, alibart2013pattern, chen2016design}. In this study, we extend (analog) neuromorphic computing to binary RRAMs, so that data storage, arithmetic, and neuromorphic computing can be performed on a single fabric. This versatility, in turn, allows the functional tiles to be readily reconfigured to compute different tasks optimally. Moreover, using binary devices for neural computing offers several advantages over analog devices. For example, the digital binary synaptic weights can be stored more reliably. The high ON/OFF ratio of binary devices helps improve the reliability and power efficiency of the system.

\begin{figure}[!t]
\vspace{-7pt}
\begin{centering}
\subfloat[]{\includegraphics[width=0.37\columnwidth]{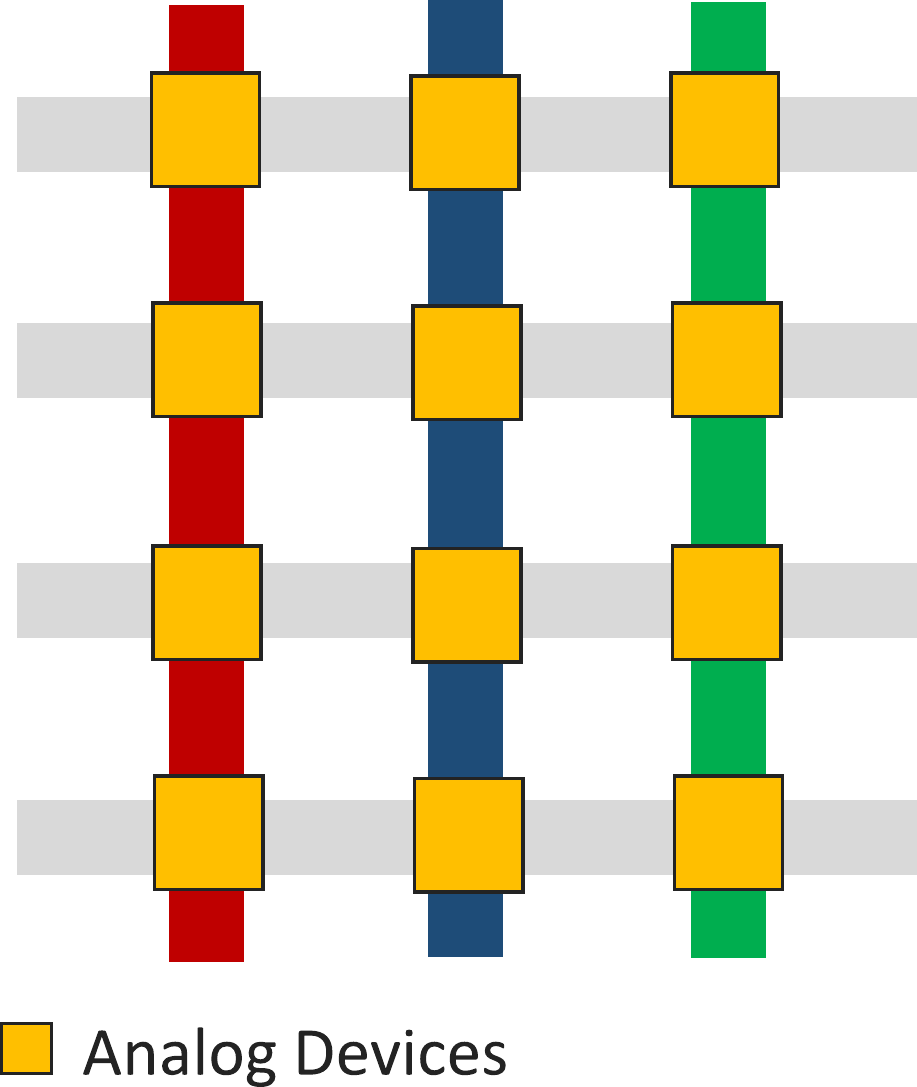}}\,\,\,\,\,\,\,\,\,\,
\subfloat[]{\includegraphics[width=0.37\columnwidth]{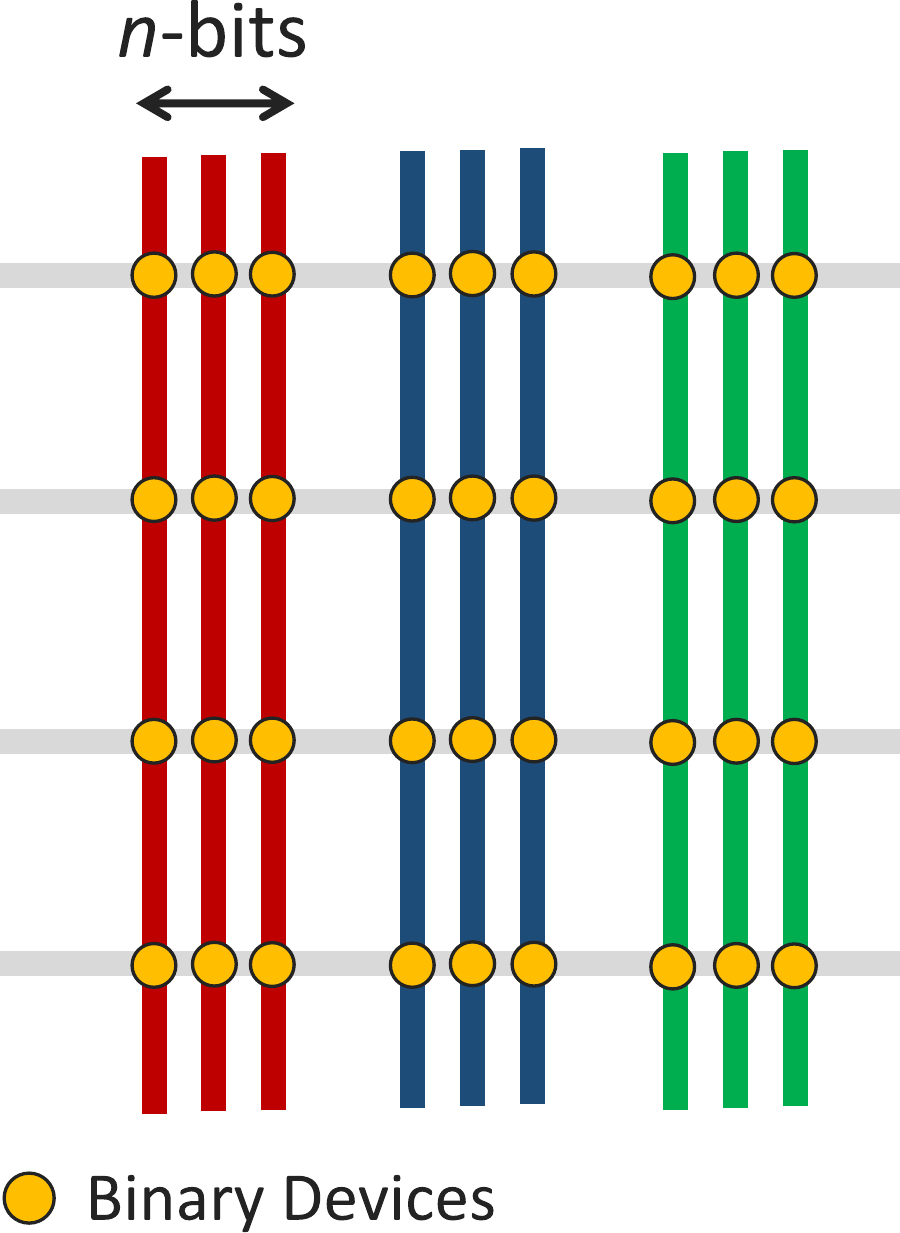}}
\par
\end{centering}
\caption{(a) Multilevel versus (b) binary coded neural networks.\label{fig_net}}
\vspace{-10pt}
\end{figure}

To map neuromorphic computing onto binary RRAM devices, we propose to encode synaptic weights in an n-bit binary representation and store a weight on n devices rather than a single analog device. Since the word length of weights used in neuromorphic computing can be quantized to just a few bits in many applications, n can be kept relatively small. In our proposed BCNN approach, each column in an analog network is replaced by n-columns in the crossbar, as shown in Figure~\ref{fig_net}. In this case, each neuron will be connected through n-columns rather a single one, where these columns are equivalent to one analog column.

The concept of using crossbar structure in neural computing is based on its native ability to sum the currents passing through a given column of synapses, weighed by the conductance values of the memristive devices, and supply the summed current to the column's (postsynaptic) neuron. This process is equivalent to an analog dot product operation of the input vector (represented by voltage pulses) and the weight vector (represented by stored conductance values). The same basic concept applies to the proposed BCNN. For example, in the case of representing each synaptic weight with n-bits, each neuron will be connected to ``n'' columns rather than one. The output current of each of the n (e.g. 4) columns represents the summation of the input current multiplied by the binary weights of this column. The equivalent analog dot product is then obtained by a binary-scaled summation of the four columns of output. Here each column output is digitized before scaling and the final sum. Analog-to-digital converters (ADCs) and adders are needed for implementing a digital neuron. We note the same components are also shared by the other two FPCA operations, namely digital computing and data storage. Since all three functions use the same devices and circuit interface, building a heterogeneous computing system using the same substrate and circuits becomes feasible.

\subsection{Analog Image Compression}
To verify the proposed concept, we performed analog image compression using the BCNN implemented on an M-Core structure. We start by training the network with a set of training images using Oja's rule and a winner-take-all (WTA) scheme~\cite{sheridan2015feature}, such that only weights associated with the winning postsynaptic neuron get updated as,
\begin{equation}
\Delta w=w_{i+1}-w_{i}=\delta y_{i}\left(x_{i}-w_{i}y_{i}\right)
\end{equation}
where ``$\Delta w$'' is update in the synaptic weights between instances ``$i$'' and ``$i+1$'', ``$\delta$'' is the learning rate, ``$x_i$'' is the presynaptic neuron input, and ``$y_i$'' is the activity of the winning postsynaptic neuron. The product ``$w_i y_i$'' value is the propagation of the winner postsynaptic response towards the presynaptic neurons. Due to the binary representation of the weight, the weights are updated using an addition or subtraction process.

\begin{figure}[!b]
\vspace{-10pt}
\begin{centering}
\subfloat[]{\includegraphics[width=0.995\columnwidth]{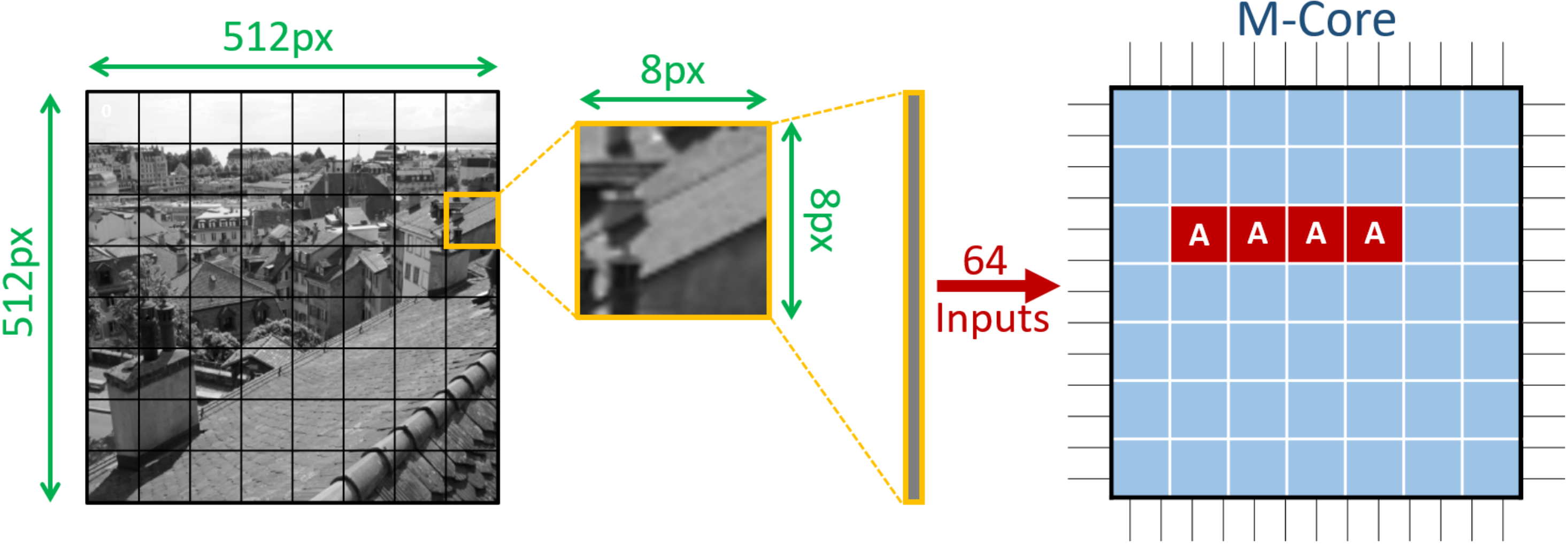}}\\
\subfloat[]{\includegraphics[width=0.47\columnwidth, height =0.235\columnwidth]{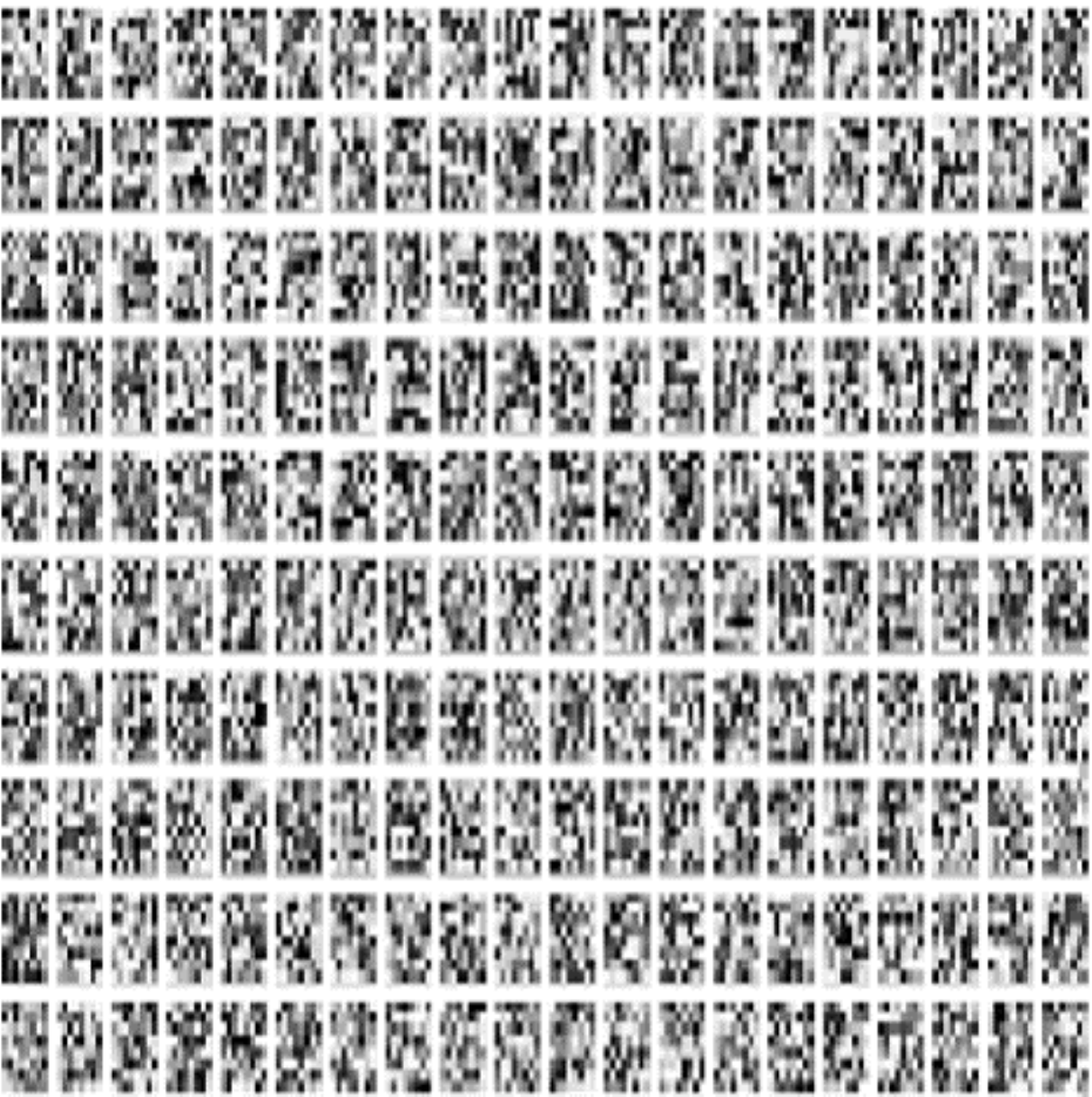}}\,\,\,\,\,
\subfloat[\label{fig:fields_b}]{\includegraphics[width=0.47\columnwidth, height =0.235\columnwidth]{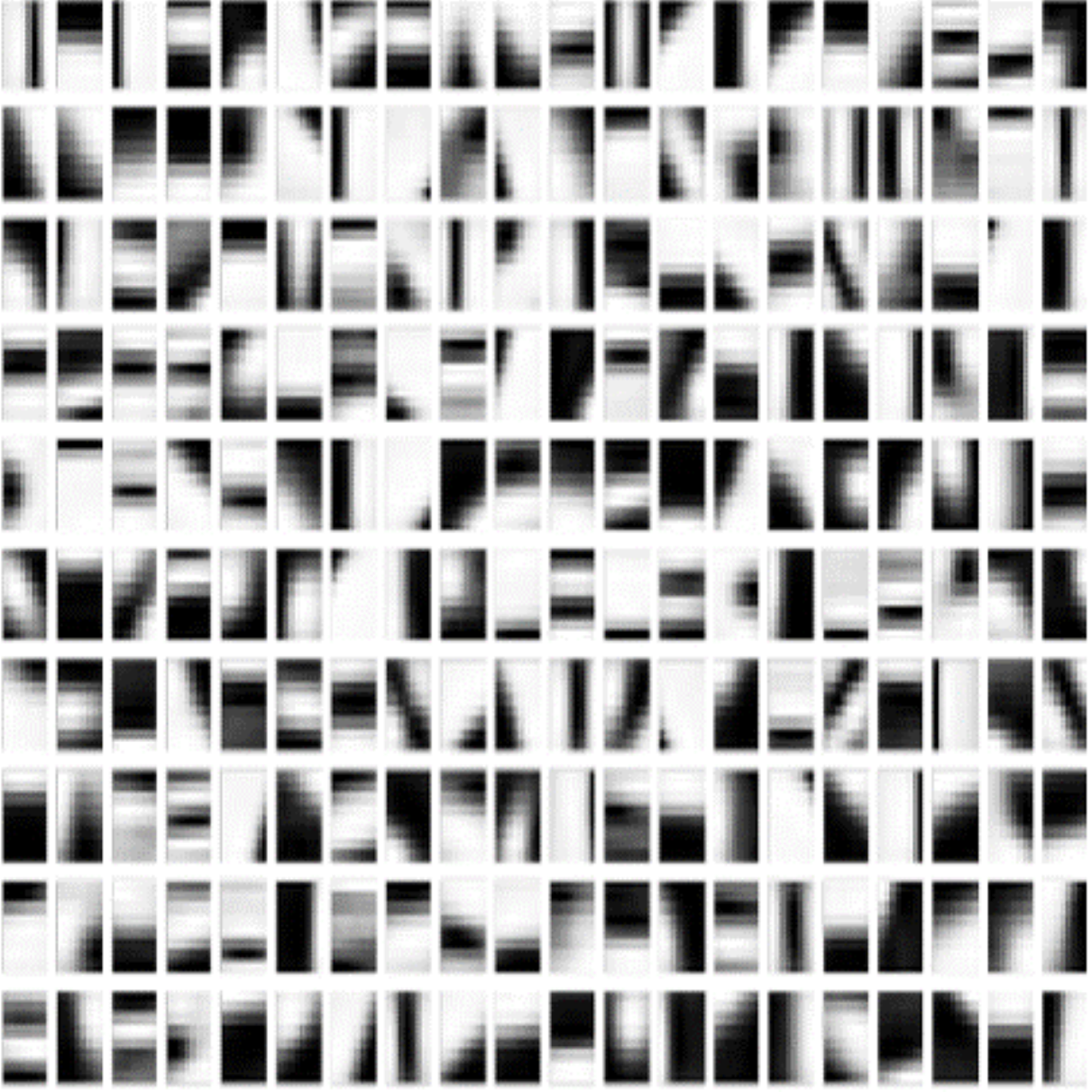}}
\par
\end{centering}
\caption{(a) A training image sliced into smaller patches, where each patch's size matches the network's input neurons. The analog tiles of an M-core are then trained with the different patches. (b, c) Two hundred dictionary elements (receptive fields) trained using the BCNN, showing (a) the original dictionary elements with random elements and (b) dictionary elements after training.\label{fig:fields}}
\end{figure}

\begin{figure*}[!t]
\vspace{-7pt}
\begin{centering}
\subfloat[Original]{\includegraphics[width=0.41\columnwidth]{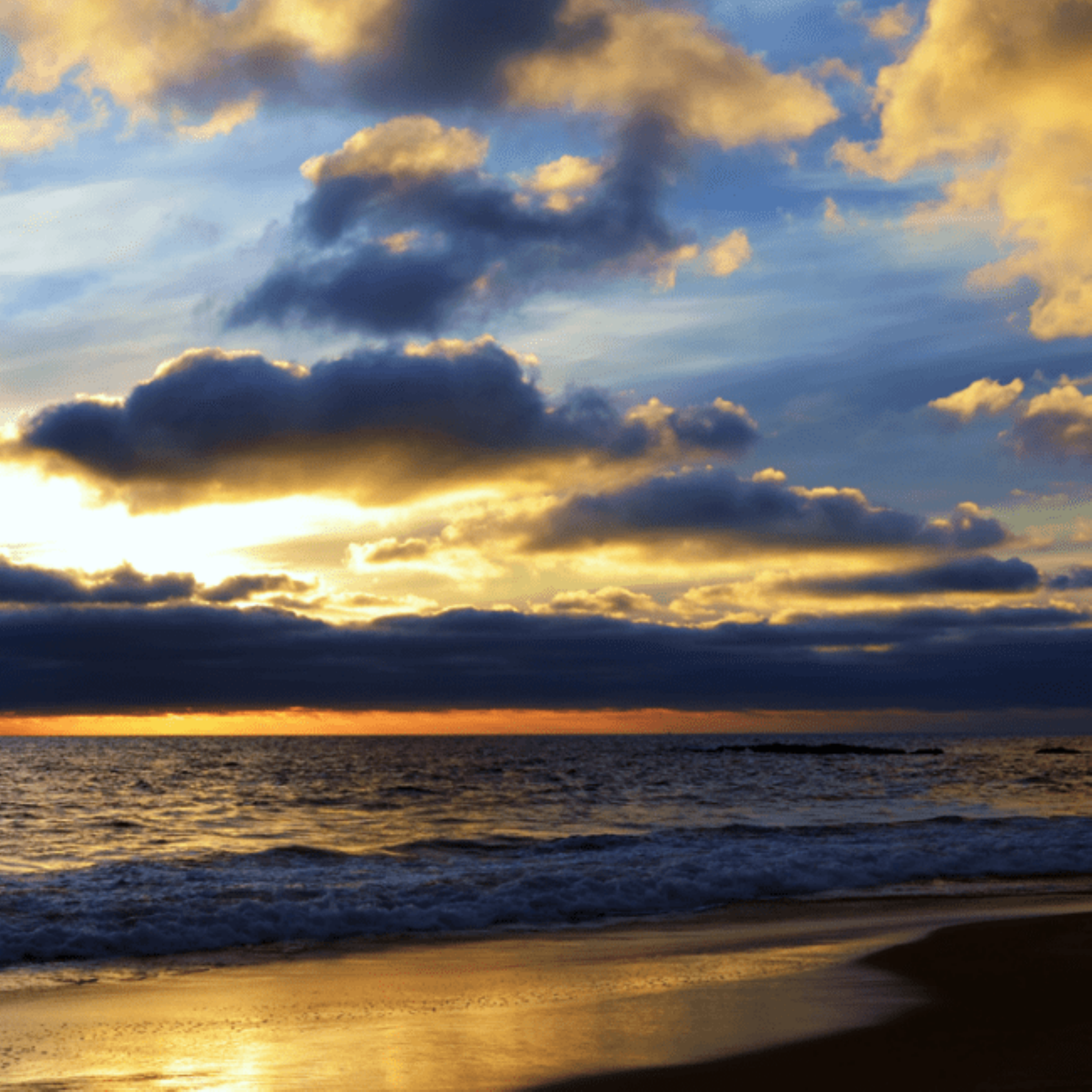}}\,\,\,\,\,
\subfloat[Reconstructed ($\lambda<0.1$)]{\includegraphics[width=0.41\columnwidth]{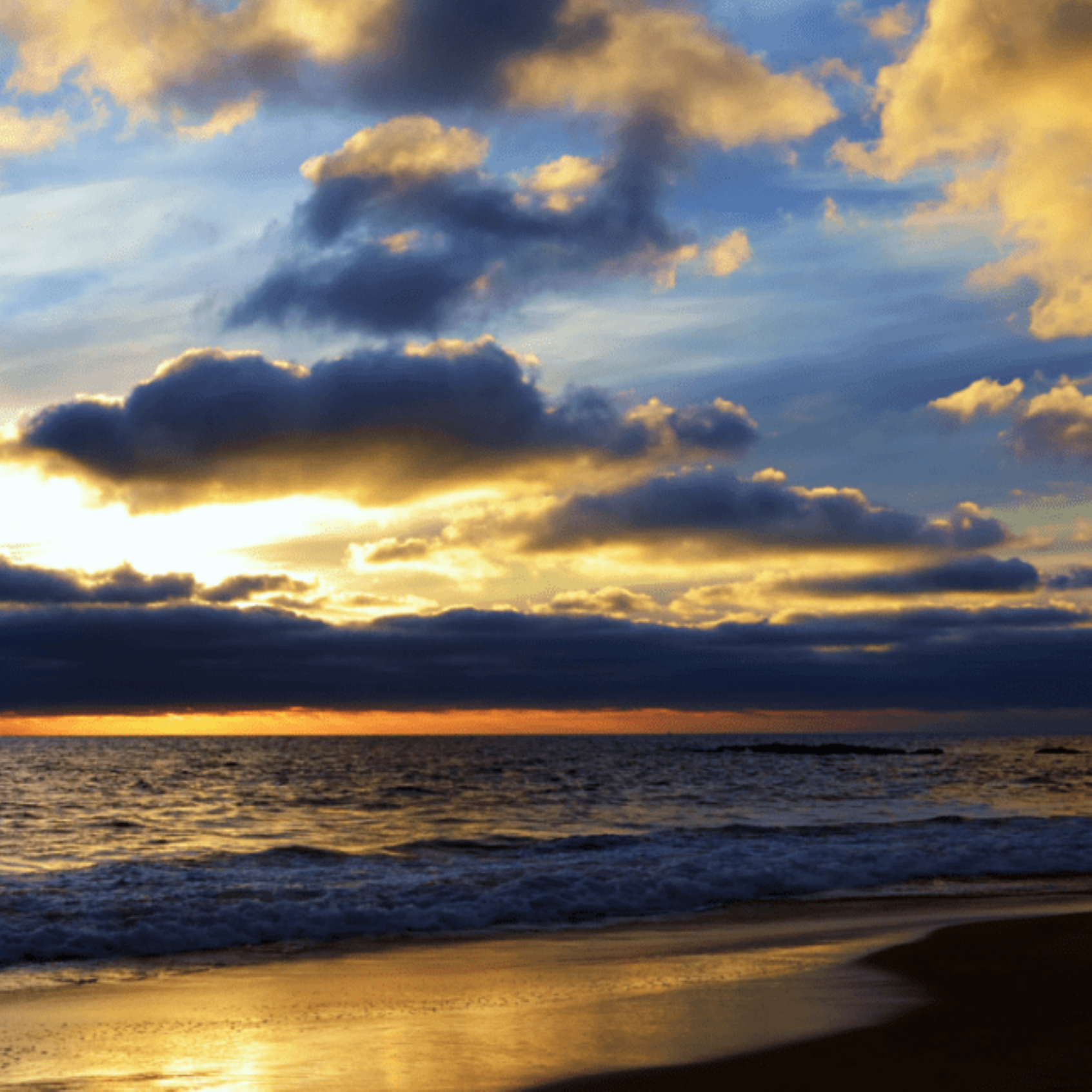}}\,\,\,\,\,
\subfloat[Reconstructed ($\lambda=0.1$)]{\includegraphics[width=0.41\columnwidth]{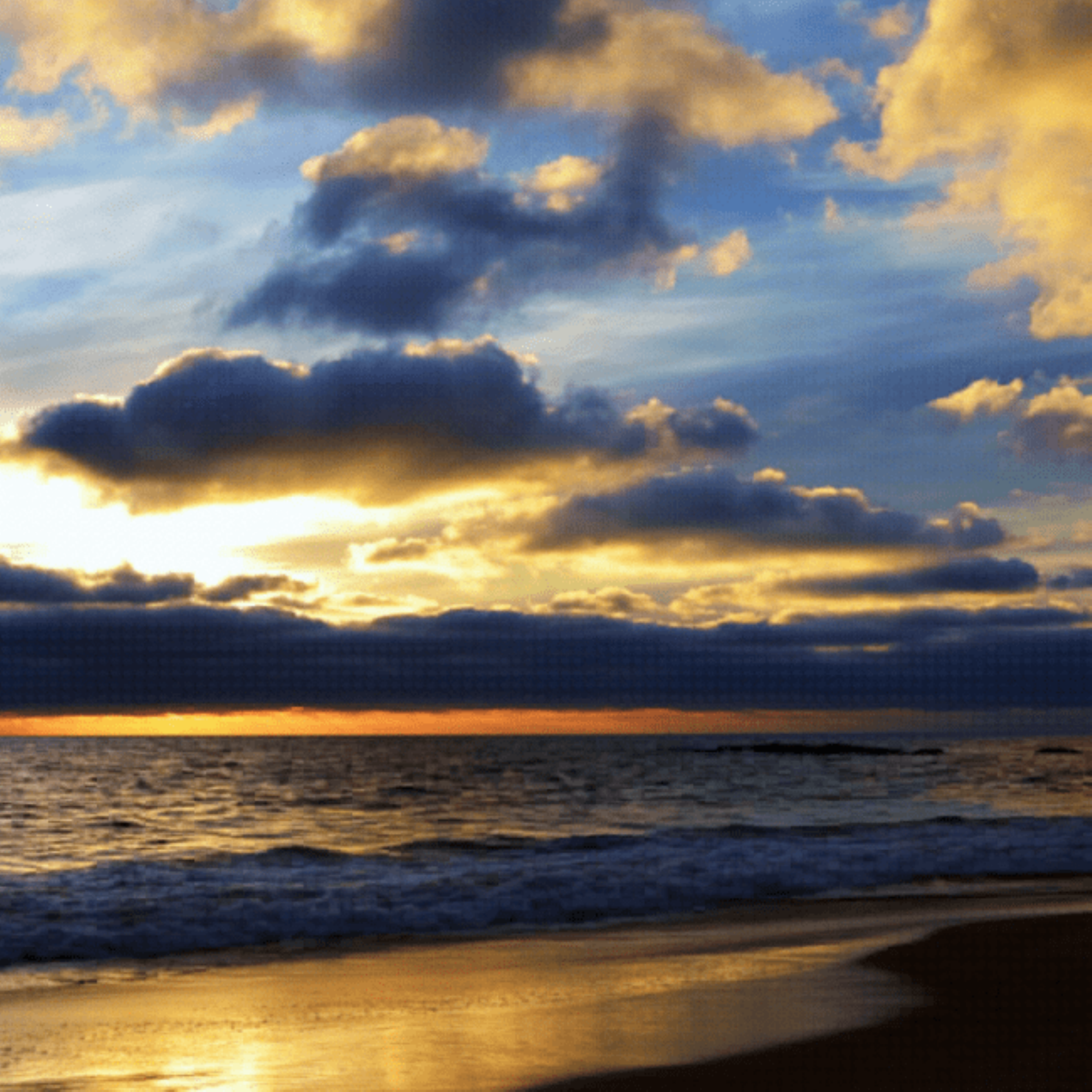}}\,\,\,\,\,
\subfloat[Reconstructed ($\lambda=0.25$)]{\includegraphics[width=0.41\columnwidth]{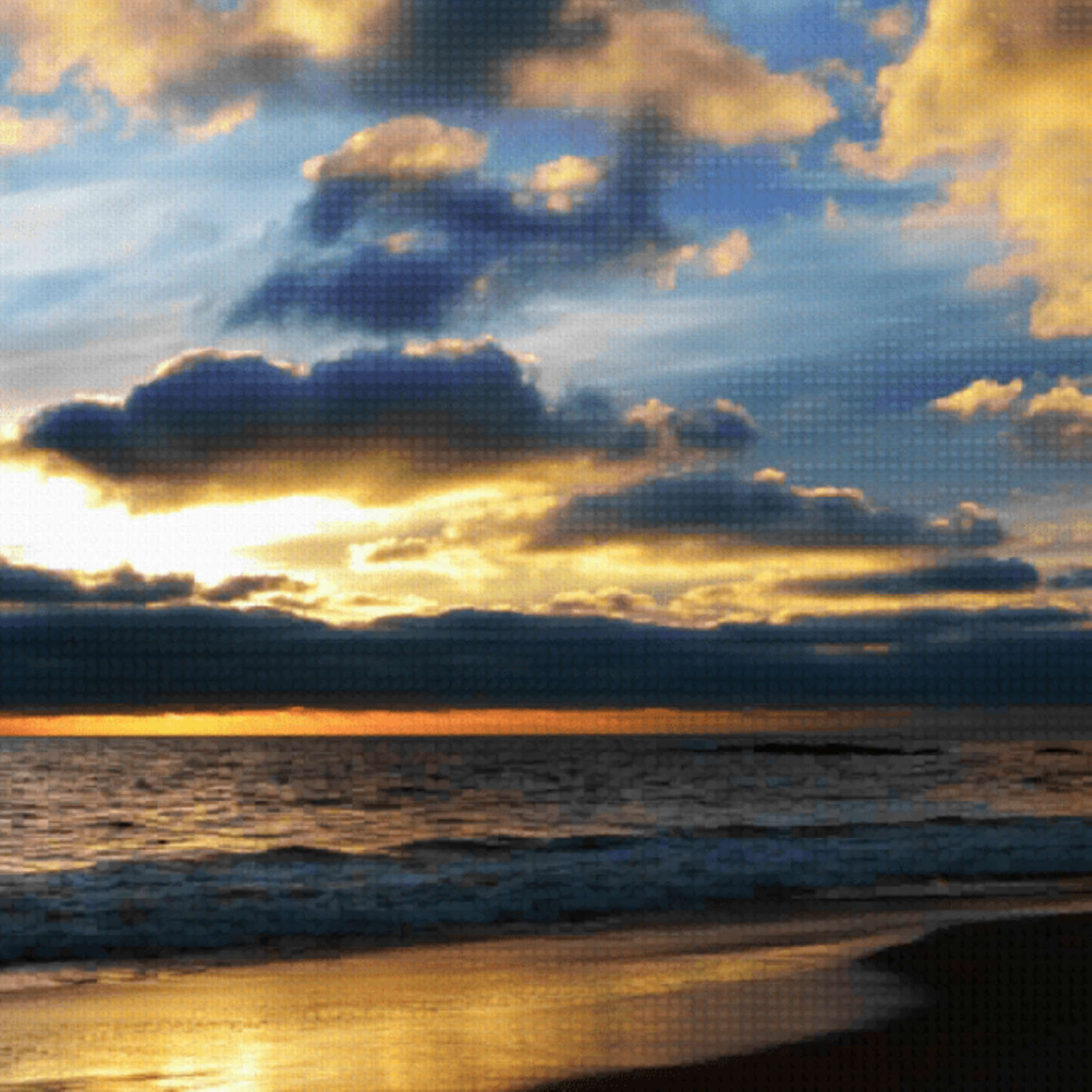}}
\par
\end{centering}
\caption{Original and reconstructed color images using the LCA algorithm implemented on the proposed binary coded neural networks.\label{fig:images}}
\vspace{-10pt}
\end{figure*}

The BCNN array is trained using a set of 37 images, each is 512$\times$512 pixels in size. The training images are sliced into 8$\times$8 pixels patches that are supplied to the network's 64 input neurons, as shown in Figure~\ref{fig:fields}a. The network in this example contains 200 dictionary elements (receptive fields), where each receptive field is represented by 16 binary columns during training, corresponding to 16-bit weights to allow incremental weight updates. After training, lower precision (e.g. 4-bit) can be used to store the trained weights at the compute/inference stage. In this case, during the training phase, more M-cores can be configured as analog resources to meet the incremental weight update requirement, then the final weights can be mapped into a system with shorter bit lengths and the stored weights can be reused many times to perform the computational tasks. Figure~\ref{fig:fields}b and c show the learned features by the network through the FPCA simulation. As expected, the trained dictionary elements resemble the receptive fields found in the biological visual cortex. It should be noted that proper training typically requires many iterations. However, training only needs to be performed once (or very infrequently), compared to the actual computational tasks.

To test the BCNN network's capability of analog image compression and reconstruction, we adopt the locally competitive algorithm (LCA)~\cite{rozell2007locally}, which is an analog sparse coding technique. The algorithm aims to reconstruct the image using the trained dictionary set, resulting in an analog compressed version of the original image while balancing sparsity (using as few neurons as possible) and accuracy constraints. The LCA algorithm can be mathematically formulated as,
\begin{equation}
u_{i+1}=u_{i}+\frac{1}{\tau}\left(\sigma_{i}-u_{i}+\eta^{T}\cdot\phi\right)
\end{equation}
where ``$u_i$'' is the membrane potential of the postsynaptic neurons at step ``$i$'', ``$\phi$'' is the matrix of the synaptic weights, ``$\tau$'' is the reconstruction time constant, ``$\sigma_{i}$'' is the neuron activation function, and ``$\eta$'' is the reconstruction error that is applied to the network as new presynaptic input:
\begin{equation}
\eta_i=x_i-\phi\cdot\sigma_{i}^{T}
\end{equation}
where ``$x_i$'' is the original presynaptic input. The two dot products ``$\eta^{T}\cdot\phi$'' and ``$\phi_i\cdot\sigma_{i}^{T}$'' are calculated by the propagation of the pre- and postsynaptic responses through the BCNN in backward and forward directions, respectively. For the neuron activity, we adopted a soft threshold function defined as,
\begin{equation}
\sigma_i=\begin{cases}
0, & \left|u_i\right|\leq0\\
4u_i-3\lambda, & 0.75\lambda<\left|u_i\right|<\lambda\\
u_i, & \left|u_i\right|>\lambda
\end{cases}
\end{equation}
where ``$\lambda$'' is the activation threshold, which in turn determines the sparsity of the reconstruction, where larger ``$\lambda$'' leads to higher compression ratio.

Figure~\ref{fig:images} shows the original and the reconstructed images using LCA implementation on BCNN with different levels of sparsity, where each synaptic weight is coded using 4 bits (implemented with four binary devices) only. We treated each of the image color channels as a separate input to the network, where each of the three color channels is reconstructed separately using the gray scale dictionaries shown in Figure~\ref{fig:fields_b}. Output from the three channels are then combined to form the reconstructed color image. We utilize the YIQ rather than the RGB color scheme to reduce intra-channel error effect to human eyes.

\begin{figure}[!b]
\vspace{-15pt}
\begin{centering}
\subfloat[\label{fig:mem_hist}]{\includegraphics[width=1\columnwidth]{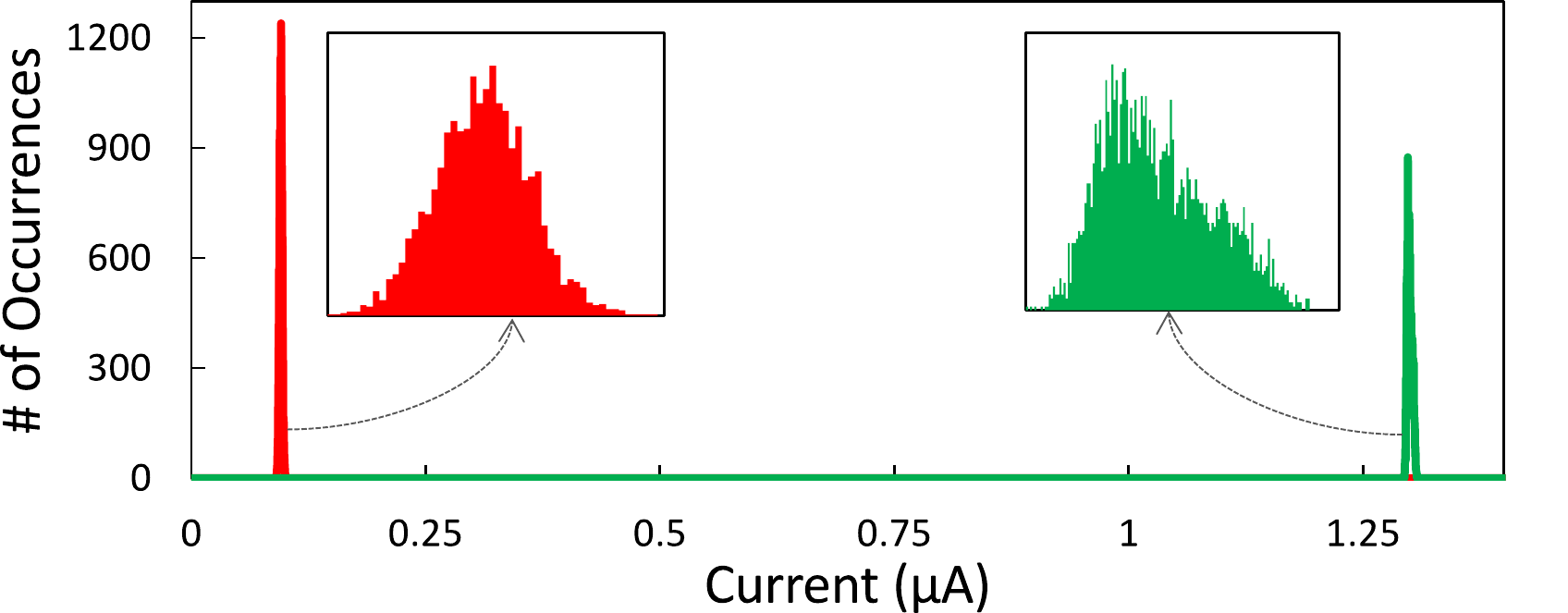}}\\
\subfloat[\label{fig:mem_scale}]{\includegraphics[width=1\columnwidth]{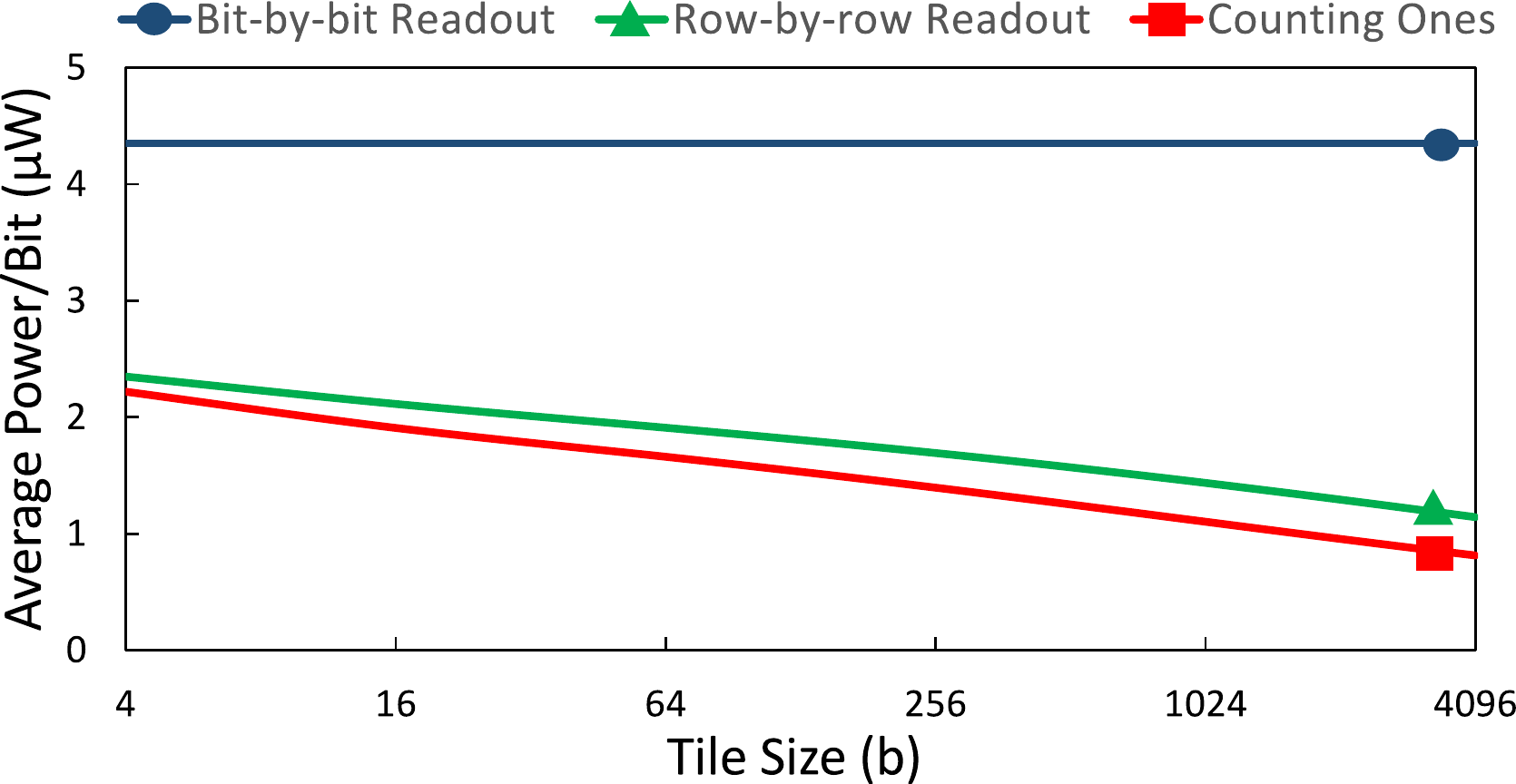}}
\par\end{centering}
\caption{(a) Readout current histogram for acceding a full row in a tile, while the rest of the M-Core is filled with random data patterns. The histogram is constructed using 32,000 simulation points. (b) Average power consumption per bit for different operations versus the tile size for a 256kb subarray.}
\end{figure}

\section{Data Storage}
Modern computing applications require high capacity and high-performance memory and storage systems. Hence, high speed, high density, and low cost per bit are the desired properties of a memory system. However, there are normally trade-offs between the goals, and current computer architecture designs are based on a memory pyramid hierarchy. At the bottom level, there is the large yet slow permanent storage, and at the top level a small and very fast cache memory and processor registers. The goal of an ideal memory hierarchy is to approach the performance of the fastest component and the cost of the cheapest one. To this end, RRAM has recently emerged as a promising candidate for future memory and storage applications. At the device level, resistive memory offers excellent scalability, fast access, low power, and wide memory margin. These attractive properties make it possible to create a simpler and flatter memory system rather than the complex pyramid memory hierarchy used today. However, a lot of RRAM's attractive features start to vanish at the system level, due to the nonidealities such as sneak paths and series line resistance that degrades the system performance.

The simplicity of the RRAM crossbar structure is also the source of its problem, namely the parasitic sneak paths~\cite{zidan2016, zidan2013memristor}. While accessing the array, current should flow only through the desired cell. However, current can sneak through other cells in the array. This parasitic current can ruin the reading and writing operations, and consumes a considerable amount of energy. Previous studies have shown that integrating binary RRAM devices with a built-in selector layer can significantly increases the nonlinearity of the device~\cite{wang2015conduction, zhou2016very}. In turn, the effect of the sneak-paths and the parasitic power consumption are decreased considerably. Such devices can also operate and switch with very low power consumption. However, the device nonlinearity do not eliminate the sneak paths interference entirely.

Most of the techniques presented in the literature to address the sneak path problem are based on the typical memory hierarchy structure, where a single cell is accessed in a sub-array at any instant of time. However, this is not the case for M-core tiles, where all the tile columns are activated at once, allowing reading an entire tile row. In this case, for a tile of size ``$n^2$'', the sneak-path interference is distributed to ``$n$'' cells rather than affecting a single cell. This improves the signal-to-noise ratio of the readout current significantly. Combining this property with RRAM devices that offer high nonlinearity will effectively eliminate the sneak-path parasitic effect. Figure~\ref{fig:mem_hist} shows the simulation results for 30k readouts from different cells in a memory core filled with 30k random data patterns. The simulation results are based on the FPCA simulation platform described earlier, and adopts the nonlinear device presented in~\cite{wang2015conduction}. The results show a large separation in the distributions of the two binary values. Such a wide separation provides sufficient memory margins to accommodate device variations.

The parallel readout not only improves the noise margins, but also reduces the energy consumption significantly. Figure~\ref{fig:mem_scale} shows the average array readout power per bit for different tile sizes. The simulation compares the classical bit-by-bit readout and the M-core based row-by-row readout. For larger tile sizes row-by-row readout saves more than 50\% of readout energy. In the same figure, we also compare the operation of counting ONEs which is the core step for arithmetic operations. Interestingly, the results show that in-memory counting using the M-Cores can be cheaper than just reading the data, which leads to an extremely fast and energy efficient arithmetic operations. It should be noted here that there is a clear dependence of the tile size on the interface circuit size, where larger tiles require larger interface area.

\begin{figure}[!b]
\vspace{-10pt}
\begin{centering}
\subfloat[\label{fig:move_a}]{\includegraphics[width=0.87\columnwidth]{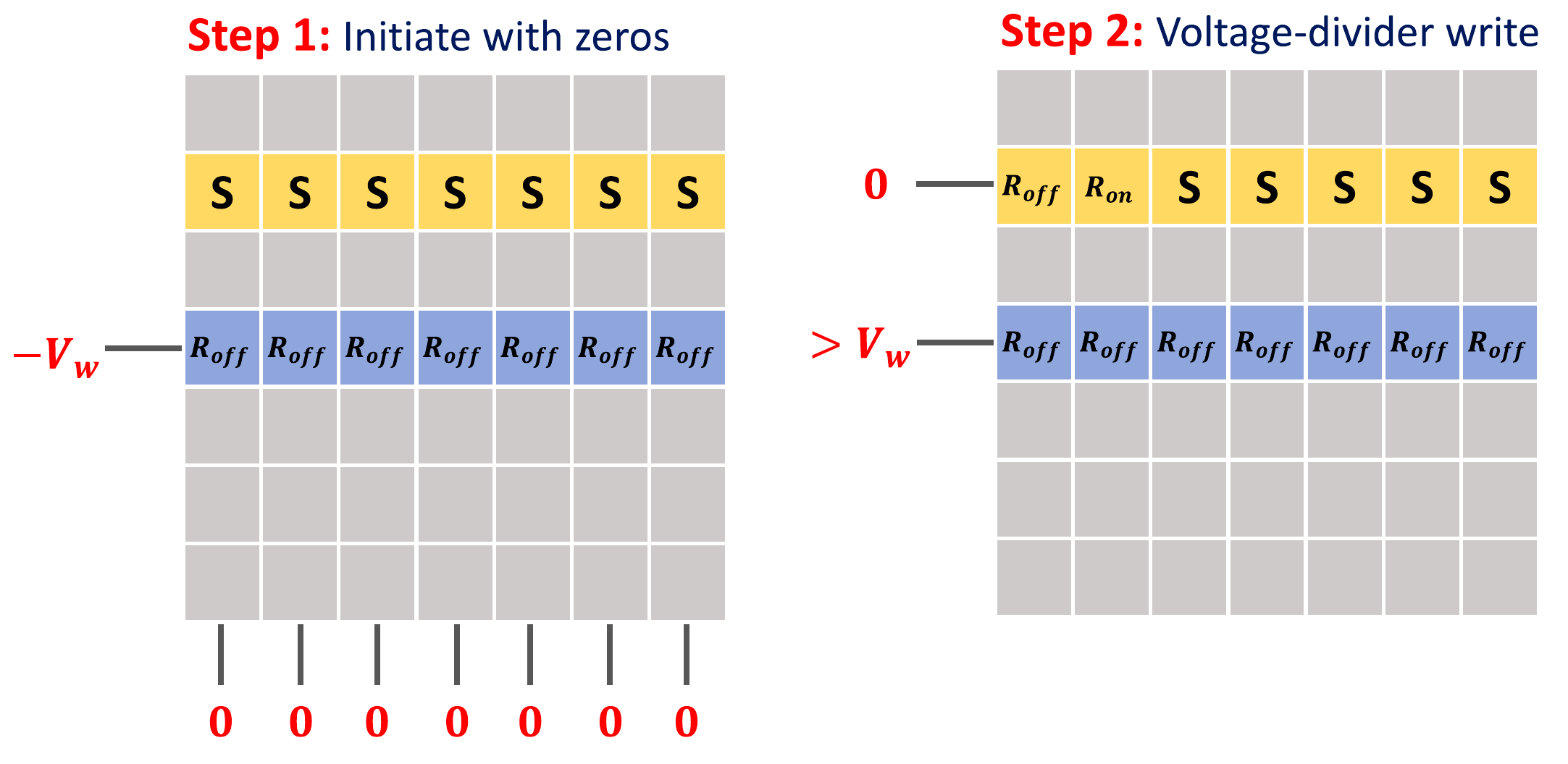}}\,\,\,
\subfloat[\label{fig:move_b}]{\includegraphics[width=0.87\columnwidth]{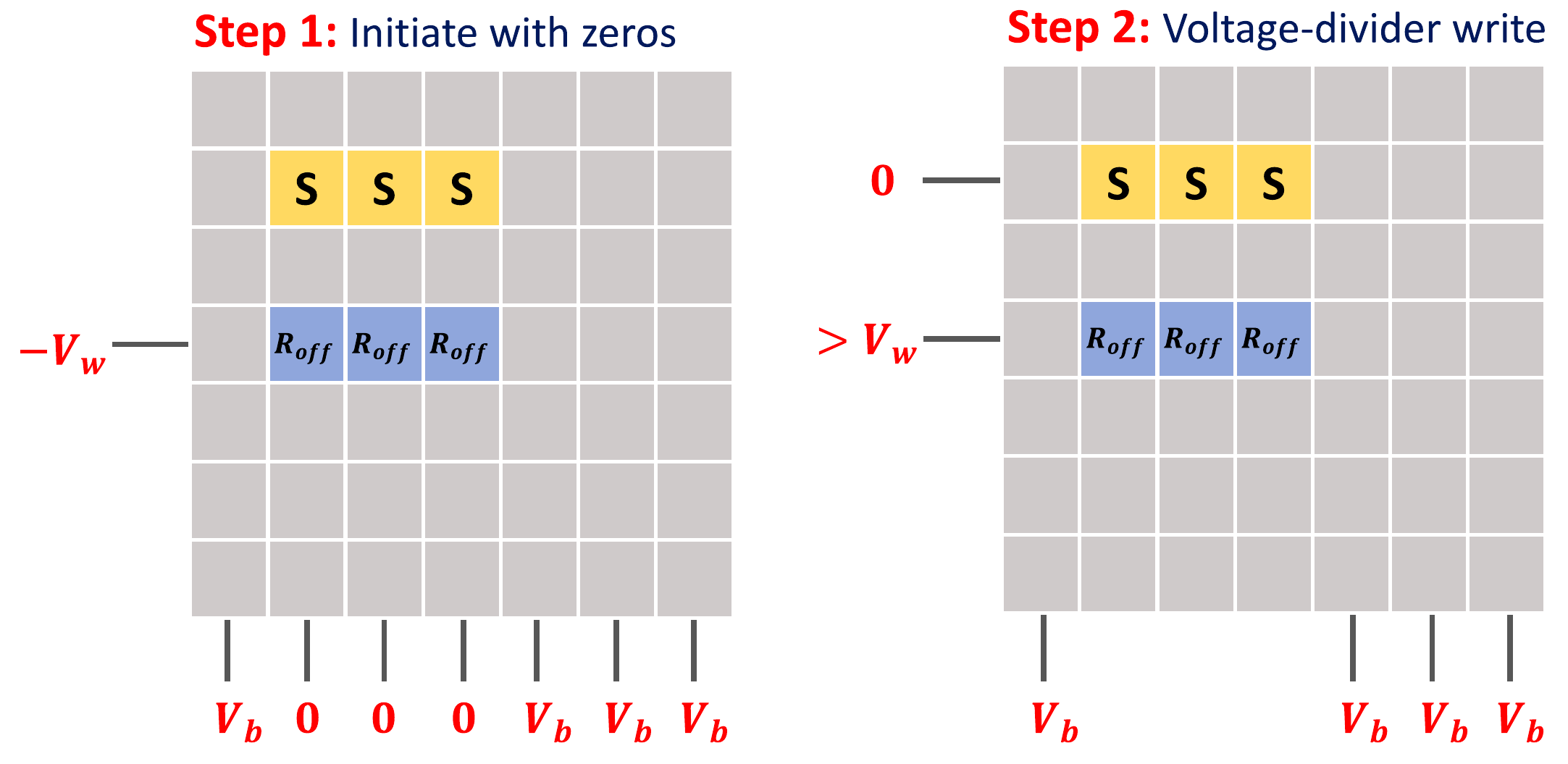}}
\par
\end{centering}
\caption{(a) Unmasked and (b) masked in-situ data shift operation, where$ V_w$ is the write threshold voltage, $V_b$ is a bias voltage, and `0' is ground. }
\end{figure}

\section{In-Situ Data Migration}
Data movement is one of the biggest challenges facing any modern computing system. The proposed architecture directly addresses the von Neumann bottleneck by effectively merging the computing and the storage units together in a single module at the physical level, and performing efficient in-memory digital and analog computing schemes. However, this does not eliminate the need for data movement completely. For example, data still need to be moved from the output from one operation to the input of the next operation, even though communication between processor and memory is no longer needed within an operation. An effective, fast technique for internal data migration based on intrinsic properties of RRAM devices is presented in this section, for efficient data migration within a tile, or between storage and computing tiles. We analyze two types of data migration. The first one is a shift movement, where data are copied either between rows or between columns. The second migration operation is the tilt movement, where data migrate between rows and columns. The two types of movements combined allow the data transfer to virtually any location in the crossbar array. The proposed data migration techniques utilize the non-linear threshold effect of RRAM devices so that properly designed voltage biasing scheme can copy from the source to the destination cells without distorting other cells in the array.

\begin{figure}[!t]
\begin{centering}
\includegraphics[width=0.97\columnwidth]{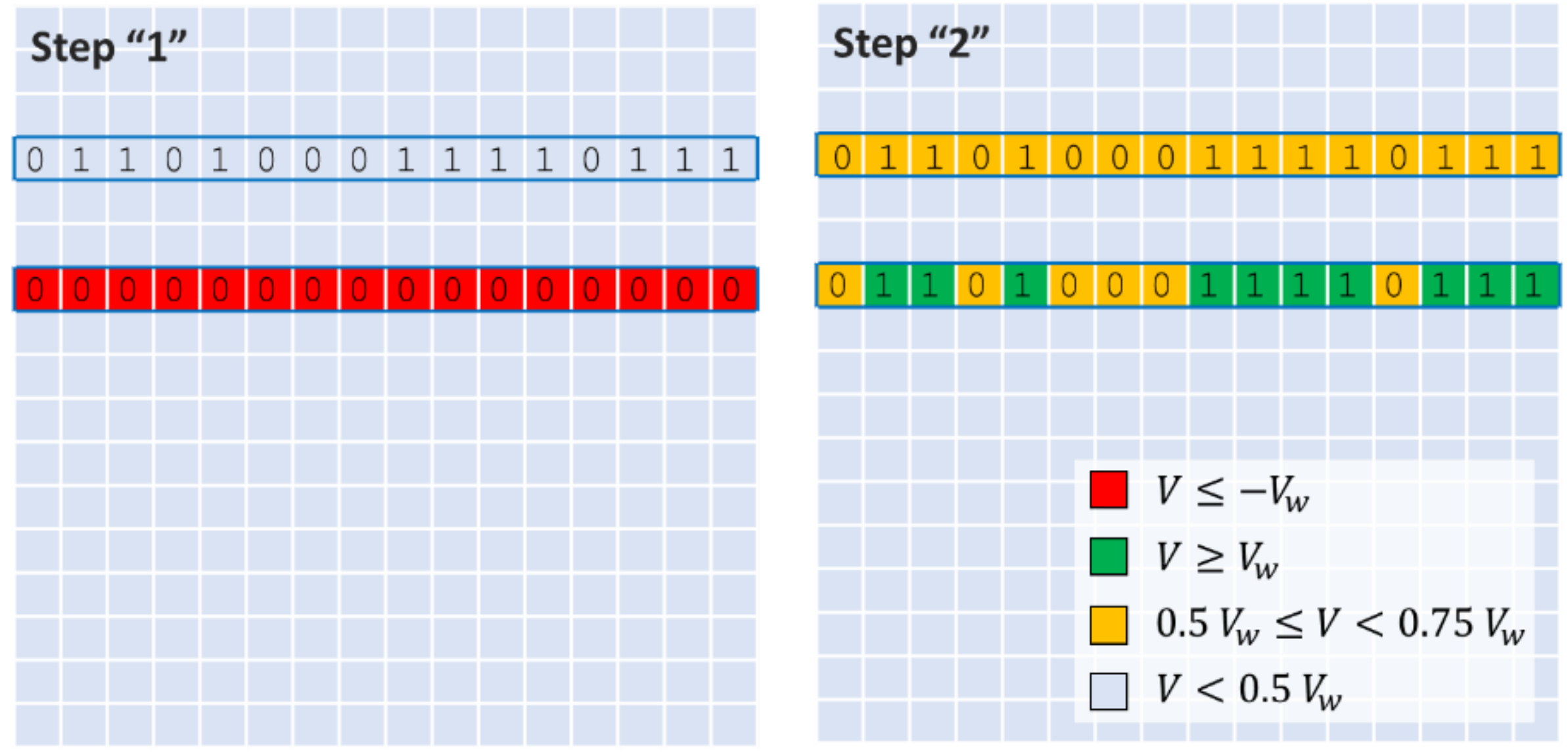}
\par\end{centering}
\caption{SPICE simulation results for the data shift operation showing the voltage drop over all the cells in an M-core tile.\label{fig:data_sim}}
\vspace{-15pt}
\end{figure}

\begin{figure*}[!t]
\begin{centering}
\subfloat[\label{fig:circuit}]{\includegraphics[height=93pt]{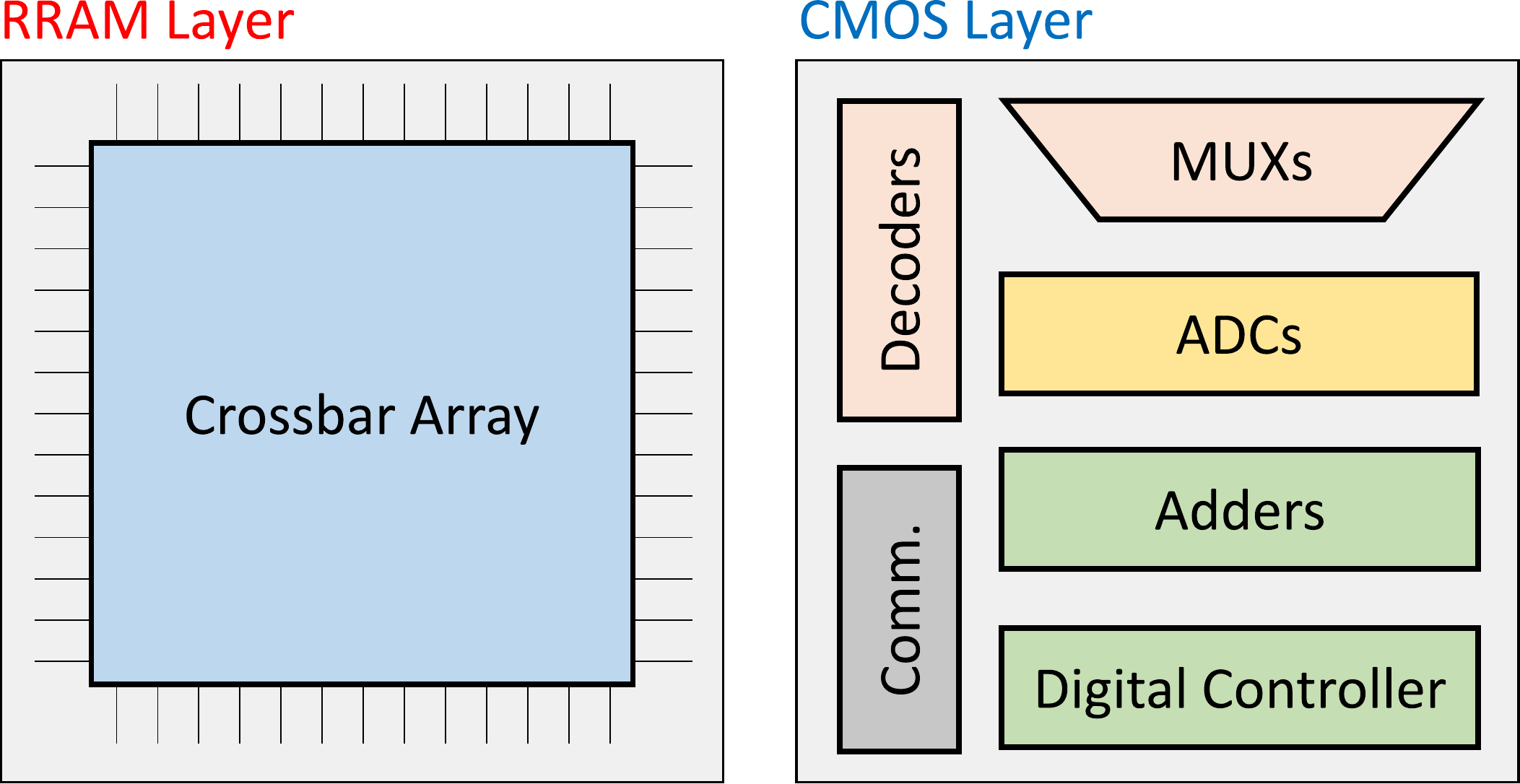}}\,\,\,\,\,\,\,\,\,\,\,
\subfloat[\label{fig:integ}]{\includegraphics[height=93pt]{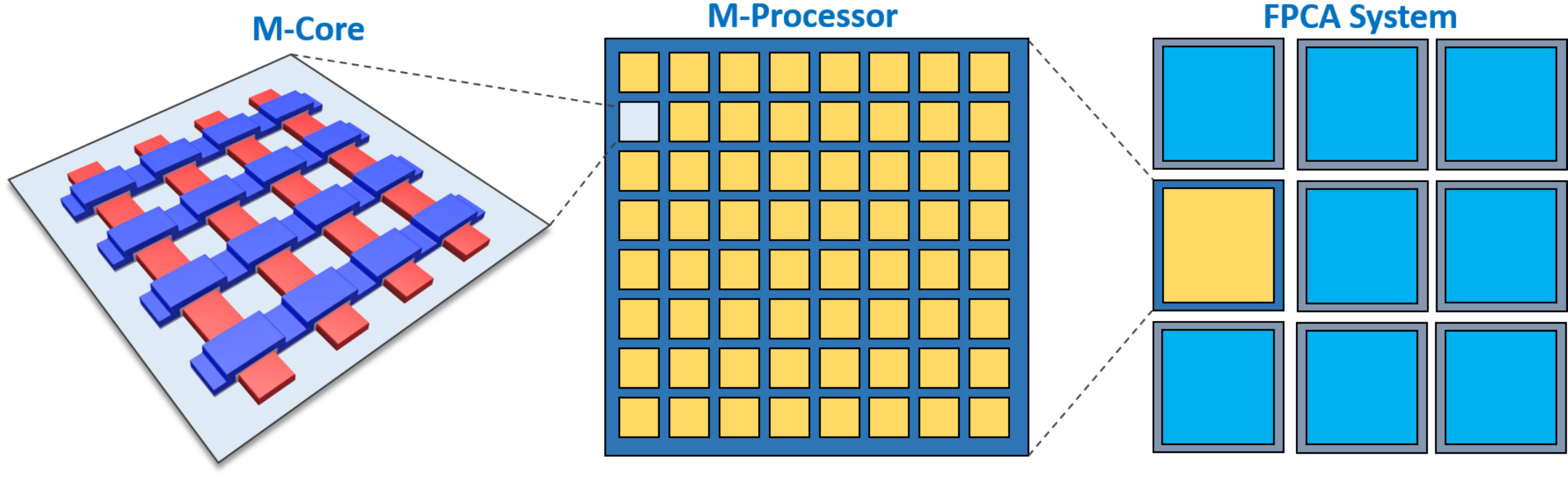}}
\par\end{centering}
\caption{(a)The content of each of the two layers of the FPCA system. (b) FPCA system hierarchy. }
\vspace{-10pt}
\end{figure*}
The data-shift method is performed in two stages as shown in Figure~\ref{fig:move_a}. The first step is to reset the destination cells to high resistance state, where ZEROs are represented by high resistance ($R_{off}$) and ONEs are represented by low resistance ($R_{on}$). In the second step, a proper voltage (e.g. 1.25x the write threshold) is applied across the source and destination rows only. This will create a voltage divider effect between the cells. In the case of the source cell storing zero ($R_{off}$), the voltage will divide equally between the source and the destination and it causes no writes to occur since the voltage across the destination cell is below the write threshold. In the other case of the source cell stores ONE, which is a low resistance state, almost all the voltage will drop over the destination cell and switch it to the low resistance state. After switching, the voltage drop is distributed equally over the two cells causing no more change to the state. Each source and destination cells in the same column (or row) will form a voltage divider pair. For a partial row (or column) migration, a masked version of the shift operation is utilized as shown in Figure~\ref{fig:move_b}. In the masked shift, a bias voltage is applied to the unselected cells forcing the voltage drop over them to be below the write threshold. This will prevent any data migration through the masked (unselected) cells.

To verify the proposed concept, a data shift operation is simulated using the FPCA simulation platform discussed earlier and the device presented in~\cite{wang2015conduction}. Figure~\ref{fig:data_sim} shows the simulation results for the designed shift process. In step one, only the desired row will have enough voltage to reset its state. All the other cells in the tile will experience a voltage drop below half the write threshold. In the second step, the voltage divider between the source and destination cells forces some destination cells to the set state based on the source cells' values. The simulation results show that the other cells in the source and destination rows will experience a safe voltage drop below three-quarters of the write threshold. Similar to data shift, the tilt operation follows the same biasing concept utilized in the data shift operations with a modified interface circuitry to support data transpose operations. It should be noted that the proposed migration process does not include any data readouts, and hence, we do not have to know the value of the cells being moved.

\section{System Integration}
\subsection{Common Interface Circuitry}
M-cores rely on two types of circuitry that are physically stacked over each other, as shown in Figure~\ref{fig:3d}. The top layer is the RRAM crossbar, which provides the system with computational and storage functions. In a typical memory application, RRAM can be constructed in the same way as a DRAM structure that is made up of subarrays, arrays, etc., to reduce capacitive loading and access delays. Similarly, an FPCA is a many-core system where the maximum continuous RRAM structure is expected to be on the order of 1 MByte acting as an M-core, whereas each M-core can be further divided into multiple (identical) crossbar sub-arrays. Each of the M-cores needs periphery circuits as decoders, MUXs, ADCs, and DACs, which are built beneath the RRAM array in the CMOS layer. The M-core can be reconfigurably divided into many tiles. Each tile is a virtual container, which is smaller than the sub-array physical size. Typically, a tile is around 32x32 or 64x64 to perform a single storage, arithmetic, or neuromorphic operation.

The decoders and the MUXs are essential for the random access operation of the RRAM layer, while the DACs and ADCs are required for sampling of the crossbar input and output signals. The CMOS layer also hosts some digital circuitry used for control and simple processing operations. Moreover, a centralized control circuitry may be needed to facilitate the overall system operation. Core-to-core data communications will be performed in the CMOS layer. It should be noted here that one of the main merits of the FPCA system is its in-memory data processing that reduces data communications significantly, and in turn reduces the interconnect circuitry complexity and area. Figure~\ref{fig:circuit} shows the set of circuitry each of the FPCA layers contains. Taking advantage of the monolithic fabrication of the system, the two layers can be connected through very high-density inter-layer vias (ILV).

To enable the different modes of operations of an M-core, a common interface circuitry that can support storage, digital and analog computing is a necessity. From the storage point of view, a reliable readout circuit for RRAM is made of ADCs and digital adders rather than a 2-bit comparator~\cite{zidan2014memristor}. The same interface circuitry can be utilized for digital computing, where the number of bits of the ADCs is determined by the virtual tile size. Larger tiles require more ADC bits but allow a higher degree of parallelism. Luckily, the BCNN digital neurons can adopt the same ADC/Adder interface. The digital neuron samples the current output and performs the leaky integrate operation using the digital adders. In addition, BCNN requires DACs to convert the native system binary data to analog inputs for the neural network. It is worth mentioning that many ADCs contain DACs within their circuitry, which eliminates the need for separate DACs. An important consideration is that the CMOS layer area should be restricted to the same order of the RRAM layer area, otherwise, the effective density of the RRAM crossbar will diminish. On the other hand, a CMOS area can be utilized by multiple interface circuitry to facilitate accessing multiple tiles per M-core concurrently for a higher throughput. To gain some insights into the CMOS layer requirements, we analyzed the ADCs, which are the largest interface units. For instance, in the case of utilizing $50nm$ RRAM feature size, each 1MB M-core is expected to occupy an area of $0.084 mm^2$. A state-of-the-art 40nm 6-bit ADC~\cite{ADC_new_1} occupies $580 \mu m^2$, which is equivalent to $0.7\%$ of a single M-core crossbar area. 64 of such 6-bit ADCs will occupy 45\% of the underneath CMOS layer, and is sufficient for counting the ONEs in a fully active 64$\times$64 tile in a parallel fashion. However, in the case of analog neuromorphic computing, the 6-bit ADC can only handle 8 rows (consuming 3-bits of the ADC) and a multi-level input of 8 states (another 3-bits). The 64 rows of the tiles can then be activated in a time multiplexed fashion in 8 time steps. The effective states of the analog input can also be increased with the aid of time multiplexing, if needed. The time multiplexing requirements are expected to be reduced or eliminated through ADC technology scaling. Other components such as DACs needed for neuromorphic computing typically consume much smaller areas compared to ADCs~\cite{dac}. The remaining CMOS layer components, including the digital adder and MUXs, usually occupy a negligible area compared to the other analog components. Finally, it is worth mentioning that recent research shows the feasibility of RRAM-based MUXs and Decoders~\cite{vontobel2009writing}, which in this case, can be built in the RRAM layer rather than in the CMOS layer.

\subsection{System Scaling}
The proposed FPCA architecture relies on medium-sized (e.g. 1MB) M-cores to provide the computing power for the system. Hence, a full system is composed of thousands of M-cores. Here arises a major challenge in how the vast number of cores will be connected together. Although in-memory data processing significantly reduces the required amount of data communications, keeping a full connectivity among all the cores is still challenging and can limit the system scaling. Here we propose two levels of hierarchy to enable a modular and scalable FPCA computing system: with a dense, locally connected structure at the lower level and a loosely connected structure at the higher level, as shown in Figure~\ref{fig:integ}. The lower hierarchical level is the M-processor, which is made of an array of fully connected M-cores. From a functional point of view, an M-processor is a digitally interfaced computing unit. Internally, the M-processor distributes the workload on analog or digital configured cores/tiles based on the workload's nature. Hence, looking from outside, an M-processor is seen as a digital processing/memory unit, while internally the computations are performed in both analog and digital domains.

At the top hierarchical level, the FPCA system is made of many of the digitally interfaced M-processors with low communication rate between them. The different levels of data communication rates are a result of the locality property of the data, where nearby M-cores, within the same M-processor, need to communicate more frequently than cores belonging to different processors. It should be noted here that, the two-level processor hierarchy is also utilized in GPU systems to manage their enormous number of tiny cores, where each set of cores are grouped in a multiprocessor unit. However, GPUs employ a totally different communications scheme that suites the graphical processing nature. In our case, the two-level hierarchy facilitates both system scalability and internal data communication requirements. Designing the FPCA as a multi-processor many-core computing system also makes it easier to control and reconfigure the system.


\subsection{Performance Estimation}
\begin{figure}[!t]
\begin{centering}
\includegraphics[width=\columnwidth]{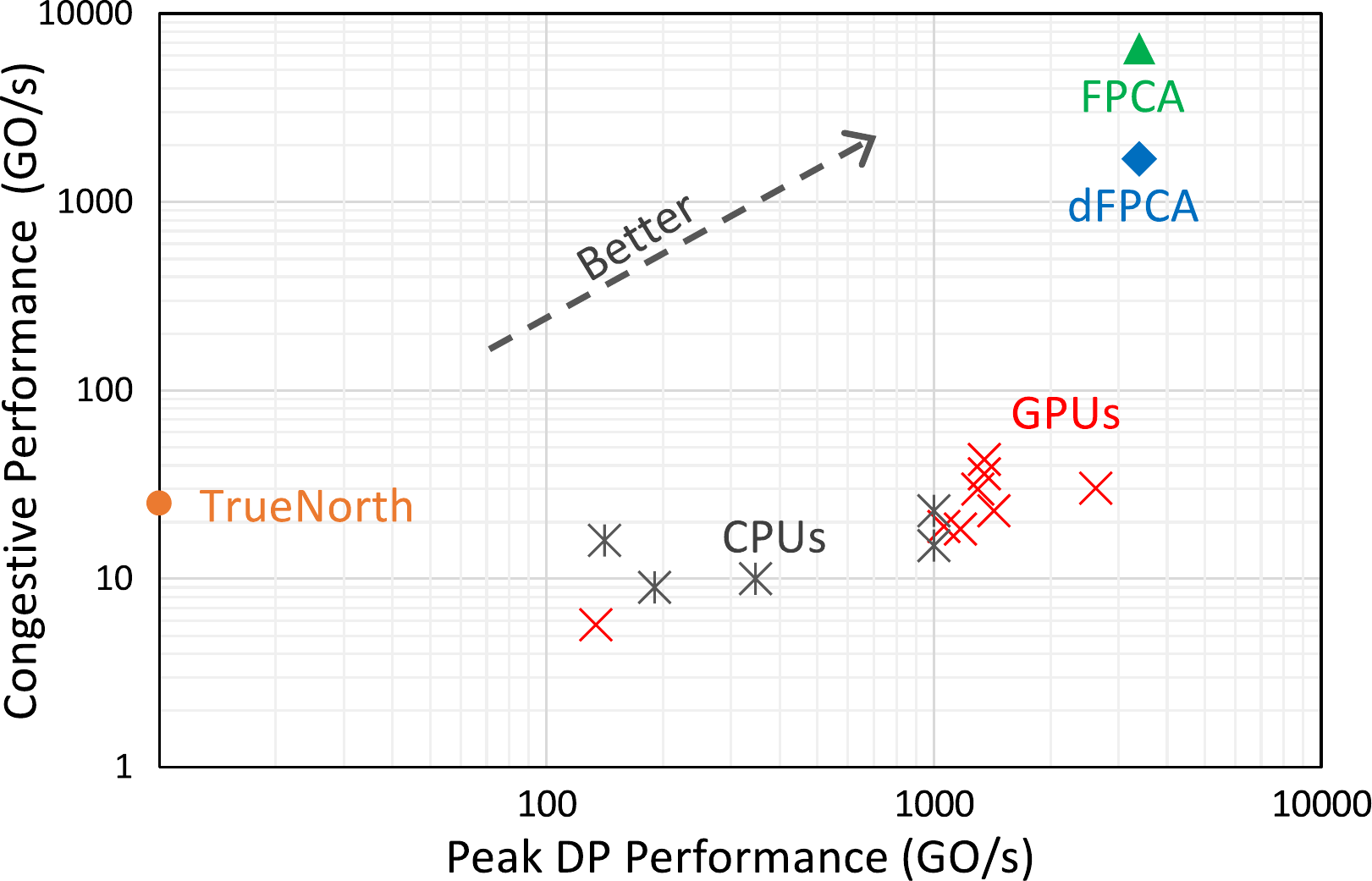}
\par\end{centering}
\caption{Classical, Neuromorphic, and FPCA computing platforms performance in Giga operations per second for traditional and congestive applications. \label{fig:performance}}
\vspace{-10pt}
\end{figure}

The widely-accepted FLOPS metric is not the optimal method to evaluate the performance of big data and cognitive applications, where memory access and matrix operations play a significant role. For many congestive applications, analog neural networks are believed to outperform classical architectures. However, benchmarking analog computing versus digital processors is still an open question. Here, we utilize a 2D performance plane to assess the FPCA performance versus classical and neuromorphic computing architectures, as shown in Figure~\ref{fig:performance}. On one axis, the peak double-precision performance is used to show the arithmetic capability of different systems, while the second axis represents the system's capability to deal with congestive problems (e.g. neuromorphic applications). Typically, conventional digital implementations of neural computing algorithms consist of successive sparse matrix-vector multiplications (SpMV). Thus, the software implementation of neural networks on a classical processor can be estimated using SpMV performance. Figure~\ref{fig:performance} shows the peak SpMV performance of various CPU and GPU implementations reported in the literature~\cite{dziekonski2011memory, tang2015optimizing, saule2013performance, kreutzer2015performance, liu2015framework, su2012clspmv, liu2013efficient, ViennaCL, bell2008efficient, yang2015performance}, where it is clearly visible that for neuromorphic and congestive applications classical processors can only achieve a small fraction of its peak FLOPS performance. This is due to many factors including the memory wall limitation, which is fundamentally addressed in the proposed FPCA system. Neuromorphic digital processors, like IBM's TrueNorth, can deliver equivalent CPU/GPU congestive performance at a significantly lower power consumption budget~\cite{merolla2014million}. On the other hand, such hardware implementations have only been used in very limited application spaces and cannot be readily reconfigured for general purpose and hard computing, e.g. arithmetic-based applications.

In order to estimate the FPCA performance, we adopted experimentally measured device and circuit data. ADCs and DACs are assumed to occupy less than 50\% of the CMOS footprint, and the whole interface circuit is designed to work at a rate of 50MHz. This rate accounts for communication delays and eases the constraints on the interface circuitry design. Applying these constraints into the system routine enables the estimation of the peak system performance for both classical and congestive applications. An FPCA system with a 8 GByte RRAM system can deliver up to 3.39 Tera double precision (DB)  operations/second, which is empowered by the natively parallel crossbar-based M-cores. However, this peak DP performance does not tell the whole story. Calculations show that for congestive applications, the FPCA system can perform SpMV operations orders of magnitude faster than both classical and digital neuromorphic architectures. For an all-digital FPCA implementation, where SpMV operations are performed using M-core arithmetic operations, the system shows 1.7 Tera DP  operation/s in congestive performance. This number increases to 6.55 Tera operation/s in the case of utilizing the analog BCNN for neuromorphic computing, after considering the time multiplexing effect. It worth mentioning here that this analog performance can be improved by using larger ADCs (thus reducing the time-multiplexing steps), but at the expense of the digital performance. Future ADCs fabricated at smaller CMOS technology nodes should further improve both the analog and digital performance. Finally, it should be noted that the system peak performance scales with the total RRAM size (i.e. total number of M-cores).

\section{Conclusion}
Continued improvements in computing power is expected to be achieved by compute- near or in memory architectures. Instead of developing accelerators based on application specific integrated circuit (ASIC) systems that need to be re-designed for each new task, the proposed FPCA system acts as a general, efficient computing fabric that can be dynamically re-configured at both the system level and the core-level to optimally perform different tasks. Based on a common physical resistive memory-centric fabric, the FPCA system can efficiently handle traditional and emerging computational tasks in a massively parallel approach. Each of the FPCA cores can be partially or fully configured to perform digital, neuromorphic, or storage operation, while largely eliminating conventional memory bottlenecks. The crossbar structure allows arithmetic operations to be performed in a natively parallel fashion that can handle concurrent vector and matrix operations. New techniques were also developed that allow the binary resistive devices to efficiently perform neuromorphic computing and in-situ data migration tasks. Altogether, the system can be tailored to achieve maximal energy efficiency based on the data flow, by dynamically allocating the basic computing fabric to storage, arithmetic, and analog including neuromorphic computing tasks. Simulations verified the potential of the proposed reconfigurable FPCA architecture to deliver orders of magnitude improvements in performance compared with conventional approaches, while offering the flexibility to satisfy general purpose computing requirements.


\section*{Acknowledgment}

The authors thank Dr. R. Dreslinski Jr. for valuable suggestions and fruitful discussions. This work was supported in part by the National Science Foundation (NSF) through grant CCF-1617315 and by the Defense Advanced Research Program Agency (DARPA) through award HR0011-13-2-0015.


%
\bibliographystyle{IEEEtran}
\bibliography{ref}


\end{document}